\documentclass[aps,prc,twocolumn,amsmath,amssymb,showpacs]{revtex4}
\usepackage{graphicx}
\usepackage{mathptmx}
\usepackage{epsfig}
\usepackage{dcolumn}
\begin{document}

\title{Level density and $\gamma$-decay properties of closed shell Pb nuclei }

\author{N.U.H.~Syed$^1$\footnote{Electronic address: n.u.h.syed@fys.uio.no}, 
M.~Guttormsen$^1$, F.~Ingebretsen$^1$, A.C.~Larsen$^1$, T.~L\"{o}nnroth$^2$, 
J.~Rekstad$^1$, A.~Schiller$^3$, S.~Siem$^1$, and A.~Voinov$^3$
\\}

\affiliation{$^1$Department of Physics, University of Oslo, P.O.Box 1048 Blindern, N-0316 Oslo, Norway\\
$^2$Department of Physics, \AA bo Akademi, FIN-20500 \AA bo, Finland\\
$^3$Department of Physics and Astronomy, Ohio University, Athens, OH-45701, USA\\}

\date{January 23, 2009}

\begin{abstract}
The level densities and $\gamma$-ray strength functions of $^{205-208}$Pb have been measured with the Oslo method, utilizing the ($^3$He, $^3$He$^{\prime}\gamma$) and ($^3$He, $\alpha$$\gamma$) reactions on the target nuclei $^{206}$Pb and $^{208}$Pb. The extracted level densities are consistent with known discrete levels at low excitation energies. The entropies and temperatures in the micro-canonical ensemble have been deduced from the experimental level density. An average entropy difference of $\Delta S  \sim 0.9 k_B$ has been observed between $^{205}$Pb and $^{206}$Pb. The $\gamma$-ray strength functions in $^{205-208}$Pb have been extracted and compared with two models; however, none of them describe the data adequately. Intermediate structures have been observed in the $\gamma$-ray strength functions for $\gamma$-ray energies below neutron threshold in all the analyzed Pb nuclei. These structures become less pronounced while moving from the doubly-magic nucleus $^{208}$Pb to $^{205}$Pb.    
\end{abstract} 


\maketitle

\section{Introduction}
Tracing average nuclear properties such as nuclear level density and $\gamma$-ray strength function in the quasi-continuum region are of particular importance. Both level density and $\gamma$-ray strength function are inputs in statistical model calculations of compound nuclear reactions and the subsequent decay of the compound system. These calculations are important for many aspects of nuclear structure studies, e.g., fission hindrance in heavy nuclei, giant resonances built on excited states, yields of evaporation residues to populate certain exotic nuclei and production of heavy elements in stellar processes. The level density is also an essential quantity for determining thermodynamic properties of nuclei, such as entropy and temperature -- quantities that describe the many-particle behavior of the system.

The level density is defined as the number of levels per unit of excitation energy. It can be obtained experimentally from different methods such as counting of resonances following neutron capture~\cite{mughabgab} and modeling of particle evaporation spectra from compound nucleus reactions~\cite{voinov0}. The $\gamma$-ray strength function characterizes the nuclear electromagnetic response. The concept of $\gamma$-ray strength function was introduced in the work of Blatt and Weisskopf~\cite{Blatt}. Most of the experimental information on the $\gamma$-ray strength function has been obtained from the study of photonuclear cross-sections for high energy $\gamma$ transitions ($E_\gamma \sim 10-20$ MeV)~\cite{dietnberman}; however, at low $\gamma$ energies experimental data are scarce.

The Oslo cyclotron group has developed a method~\cite{schl0} to isolate the first $\gamma$-rays emitted in all decay cascades at various initial excitation energies. The energy distribution of these primary (or first-generation) $\gamma$-rays provides information on the level density and the $\gamma$-ray strength function. The Oslo method has previously been applied for rare earth nuclei, obtaining information on the level density and average $\gamma$-decay properties in this mass region~\cite{melby,MGpos}. Recently, the method has been extended to other mass regions like Fe, Mo, V and Sc~\cite{schl02, Mo1,Mo2,cecil01,cecil02}. In order to check the validity of the method in cases where nuclei have low level density and strong Porter-Thomas fluctuations~\cite{PortarThomas}, the $^{27,28}$Si nuclei were studied with gratifying results~\cite{Si}. These achievements have encouraged us to apply the method on closed shell $^{205-208}$Pb nuclei, where the decay properties are less statistical due to shell effects and more dominated by single-particle selection rules.

Based on the shell model, the nucleon configuration of lead isotopes is such that the protons fill the $h_{11/2}$ shell, and the valence neutrons reside in the $i_{13/2}$ shell. The chain of $A = 205 - 208$ lead isotopes in general, and the doubly-magic $^{208}$Pb nucleus with $Z = 82$ and $N = 126$ in particular, are interesting nuclei for studying the nuclear structure and $\gamma$-decay properties at and in the vicinity of a two major shell closures.

In the following sections a brief outline of the experimental method and data analysis are given. The level densities, thermodynamic properties and the $\gamma$-ray strength functions will be discussed. Finally, a summary and conclusions are drawn.  

\section{Experimental method and data analysis}

The experiments were conducted at the Oslo Cyclotron Laboratory (OCL) using a 38-MeV $^3$He ion beam. Self-supporting targets of $^{206}$Pb and $^{208}$Pb metallic foils, enriched to 99.8$\%$ and 99.9$\%$ and with thicknesses of 4.7 mg/cm$^2$ and 1.4~mg/cm$^2$, respectively, were used. The experiments ran for twelve days with a beam current of $\approx$0.6 nA. \\

The bombardment of $^3$He ions on the Pb targets opens a number of reaction channels, such as ($^3$He, $^3$He$^\prime$$\gamma$), ($^3$He, $\alpha \gamma$), ($^3$He, $xn\alpha \gamma$) and ($^3$He, $^3$He$^\prime xn\gamma$). The following reactions are analyzed in the present study:
\begin{enumerate}
\item  $^{208}$Pb($^3$He,$^3$He$^\prime$$\gamma$)$^{208}$Pb
\item $^{208}$Pb($^3$He, $\alpha$$\gamma$)$^{207}$Pb
\item $^{206}$Pb($^3$He, $^3$He$^\prime$$\gamma$)$^{206}$Pb
\item $^{206}$Pb($^3$He, $\alpha$$\gamma$)$^{205}$Pb
\end{enumerate}

Particle-$\gamma$ coincidences were measured for $^{205-208}$Pb using the CACTUS multidetector array~\cite{Cactus}. The charged particles were detected by eight collimated $\Delta$E$ - $E type Si particle telescopes, placed at a distance of 5 cm from the target and making an angle of 45$^\circ$ with the beam line. The thicknesses of the $\Delta$E and E detectors are $\approx$145 $\mu$m and $\approx$1500 $\mu$m, respectively. For the detection of $\gamma$-rays, 28 5$^{\prime\prime}\times$5$^{\prime\prime}$ NaI detectors were used, surrounding the particle telescopes and the target. The total efficiency of the NaI detectors is $\sim$ 15$\%$ of 4$\pi$.

The data analysis consists of four main steps: (i) making the particle-$\gamma$ coincidence matrix, (ii) unfolding the total $\gamma$-ray spectra, (iii) extracting the first generation $\gamma$-ray spectra and (iv) the factorization of the first generation $\gamma$-ray matrix into level density and $\gamma$-ray transmission coefficient. From the known $Q$ values and reaction kinematics the ejectile energy can be transformed into initial excitation energy of the residual nuclei. Using the particle-$\gamma$ coincidence technique, each $\gamma$-ray can be assigned to a cascade depopulating an excitation energy region in the residual nucleus. Thus, the particle-$\gamma$ coincidence measurements give a total $\gamma$-ray spectrum for each excitation energy bin. Each row of the coincidence matrix corresponds to a certain excitation energy ($E$) while each column corresponds to a certain $\gamma$-ray energy ($E_{\gamma}$). In $^{205,206}$Pb and $^{207,208}$Pb the excitation energy bins are chosen to be 240 keV/channel and 220 keV/channel, respectively.

\begin{figure}
\centering
\includegraphics[height=0.8\textheight]{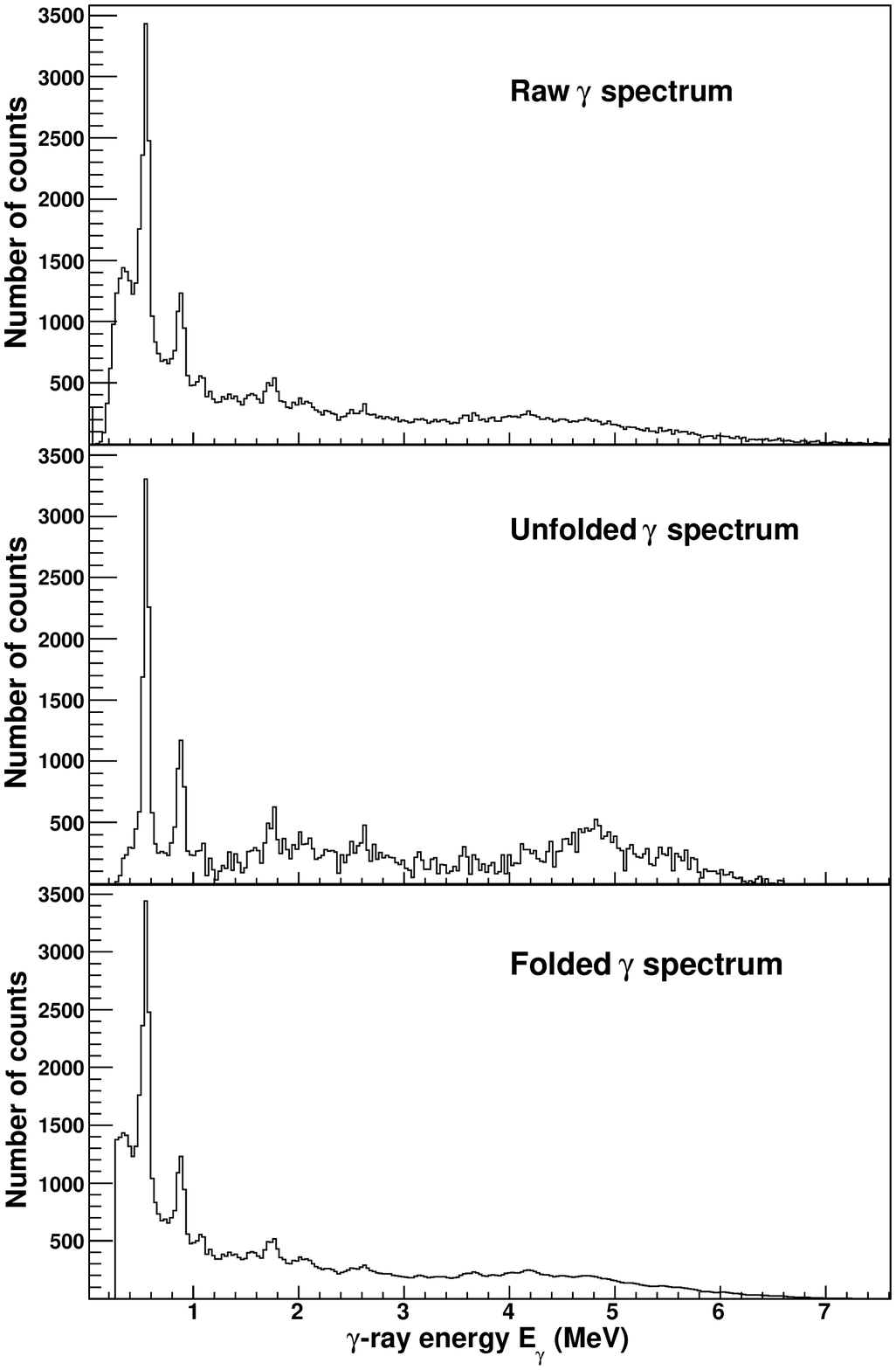}
\caption{Unfolding of the $\gamma$-ray spectrum of $^{207}$Pb measured for the excitation energy region $E = 4.5 - 6.7$ MeV. Note the similarity of the raw and folded spectrum.}
\label{fig:unfolding}
\end {figure}

The $\gamma$-ray spectra are corrected for the NaI response function. The unfolding procedure of Ref.~\cite{gutt6} is employed for this purpose, which is based on the Compton-subtracting technique that prevents additional spurious fluctuations in the unfolded spectrum. The reliability of the unfolding technique is tested for $^{207}$Pb $\gamma$-ray spectra, see Fig.~\ref{fig:unfolding}. Here, the raw $\gamma$-ray spectrum from the excitation energy region $E = 4.5 - 6.7$ MeV is compared with the spectrum that has been folded after unfolding. The good agreement with the raw $\gamma$-ray spectrum and the folded spectrum gives confidence in the employed unfolding technique. The set of unfolded $\gamma$-ray spectra are organized in a $(E, E_\gamma)$ matrix, which comprises the energy distribution of all $\gamma$-rays from all decay cascades as a function of excitation energy. This matrix of total, unfolded $\gamma$-ray spectra is the basis for the next step of the Oslo method.

The first generation (primary) $\gamma$-ray energy distribution is extracted by an iterative subtraction technique described in Ref.~\cite{gutt0}. The basic assumption of the subtraction method is that the $\gamma$-decay pattern from any excitation energy bin is independent of whether the state is populated directly via scattering or neutron pick-up, or through $\gamma$-decay from a higher-lying excited state.

The basic assumption may not be fulfilled if the direct reaction at lower excitation bins, whose $\gamma$-spectra are utilized to subtract higher-generation $\gamma$-rays from total spectra at higher excitation energies, do not favor some levels within the excitation bin that are populated by $\gamma$-rays from above. Such situations may cause that some $\gamma$-rays are not fully subtracted from the total spectrum.

The influence of a possible different selectivity of levels at one excitation energy in the direct reaction compared to $\gamma$-decay from higher lying levels, is expected to be most pronounced when only a few levels are present in the excitation bin. Thus, these considerations are increasingly important for nuclei in the vicinity of closed shells having low level density. It is therefore necessary to be cautious when applying the Oslo method to the Pb nuclei, and to carefully check that the method gives reasonable results compared to other data.

\begin{figure}
\centering
\includegraphics[height=0.8\textheight]{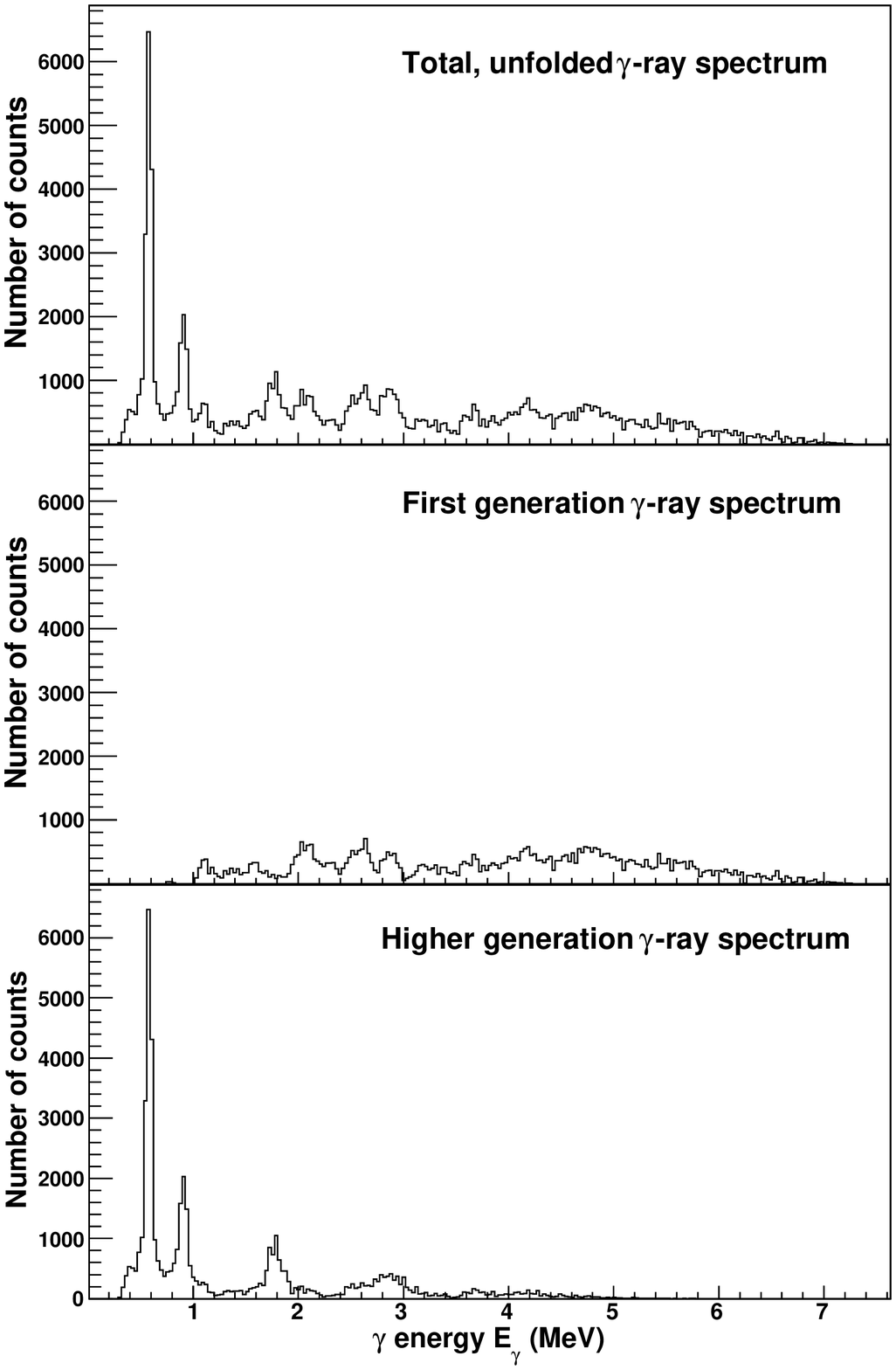}
\caption{Extraction of first generation $\gamma$-rays (middle panel) from the total, unfolded $\gamma$-ray spectrum (top panel) by subtracting the second and higher generation $\gamma$-rays (bottom panel) in $^{207}$Pb between excitation energies of 4.5 and 6.7 MeV.}
\label{fig:fg}
\end {figure}

In Fig.~\ref{fig:fg} the application of the first generation $\gamma$-rays extraction technique is shown for $^{207}$Pb. The first generation $\gamma$-rays are extracted by subtracting the weighted sum of higher generation $\gamma$-rays from the total unfolded $\gamma$-ray spectrum. Figure~\ref{fig:fg} shows the total unfolded $\gamma$-ray spectrum, the first generation $\gamma$-ray spectrum, and the second and higher generation $\gamma$-spectrum from $E = 4.5 - 6.7$ MeV for $^{207}$Pb. It is seen how the higher generation $\gamma$-rays are well separated from the total $\gamma$-ray spectrum in the specified excitation energy region.

\section{Nuclear level densities}
\subsection{Application of the Axel-Brink hypothesis}
The energy distribution of primary $\gamma$-rays emitted from a well-defined initial excitation energy provides information on the level density and the $\gamma$-strength function. First, the experimental primary $\gamma$-ray matrix $P(E, E_\gamma)$ is normalized for the various excitation energy bins $E$. This is done by summing $P(E, E_\gamma)$ over all $\gamma$-ray energies $E_\gamma$, for each excitation energy bin $E$ such that
\begin{equation} 
\sum_{E_\gamma = E_\gamma^\mathrm{min}}^{E}  P(E, E_\gamma) = 1.
\label{1}
\end{equation}
The entries of the first generation matrix are the probabilities $P(E, E_\gamma)$ that a $\gamma$-ray of energy $E_\gamma$ is emitted from excitation energy $E$. In accordance with Fermi's golden rule, the decay rate is proportional to the level density of final states and the square of the transition matrix element. Thus, entries of the primary $\gamma$-ray matrix $P(E,E_\gamma)$ in the statistical region are proportional to the level density $\rho(E-E_\gamma)$ and the $\gamma$-ray transmission coefficient $\cal {T}$$(E_\gamma)$:
\begin{equation}
P(E, E_{\gamma}) \propto  \rho (E -E_{\gamma}) {\cal{T}}  (E_{\gamma}).\
\label{eqn:2}
\end{equation}
In the factorization procedure of the two-dimensional primary $\gamma$-ray matrix, a lower limit of the excitation energy region $E \sim 3.0$ MeV was used for all the analyzed nuclei in order to exclude the main part of non-statistical transitions from our data. The normalized experimental primary $\gamma$-ray matrix $P(E,E_\gamma)$ can be approximated by
\begin{equation}
P_{\mathrm{th}}(E, E_\gamma) = \frac{\rho(E-E_\gamma) {\cal T} (E_\gamma) }{ \sum_{E_\gamma =E_\gamma^\mathrm{min}}^E \rho(E-E_\gamma) {\cal T} (E_\gamma)}. \
\label{eqn:3} 
\end{equation}
The first trial function for $\rho$ is assumed to be unity i.e., $\rho = 1$ and the corresponding $\cal {T}$ can be determined by Eq.~(\ref{eqn:3}). Then, a $\chi^2$ minimum is calculated for each data point of $\rho$ and $\cal {T}$ simultaneously. 
This procedure is repeated until a global least-square fit is achieved for all the data points of $P(E, E_\gamma)$. 

\begin{figure}
\centering
\includegraphics[height=0.8\textheight]{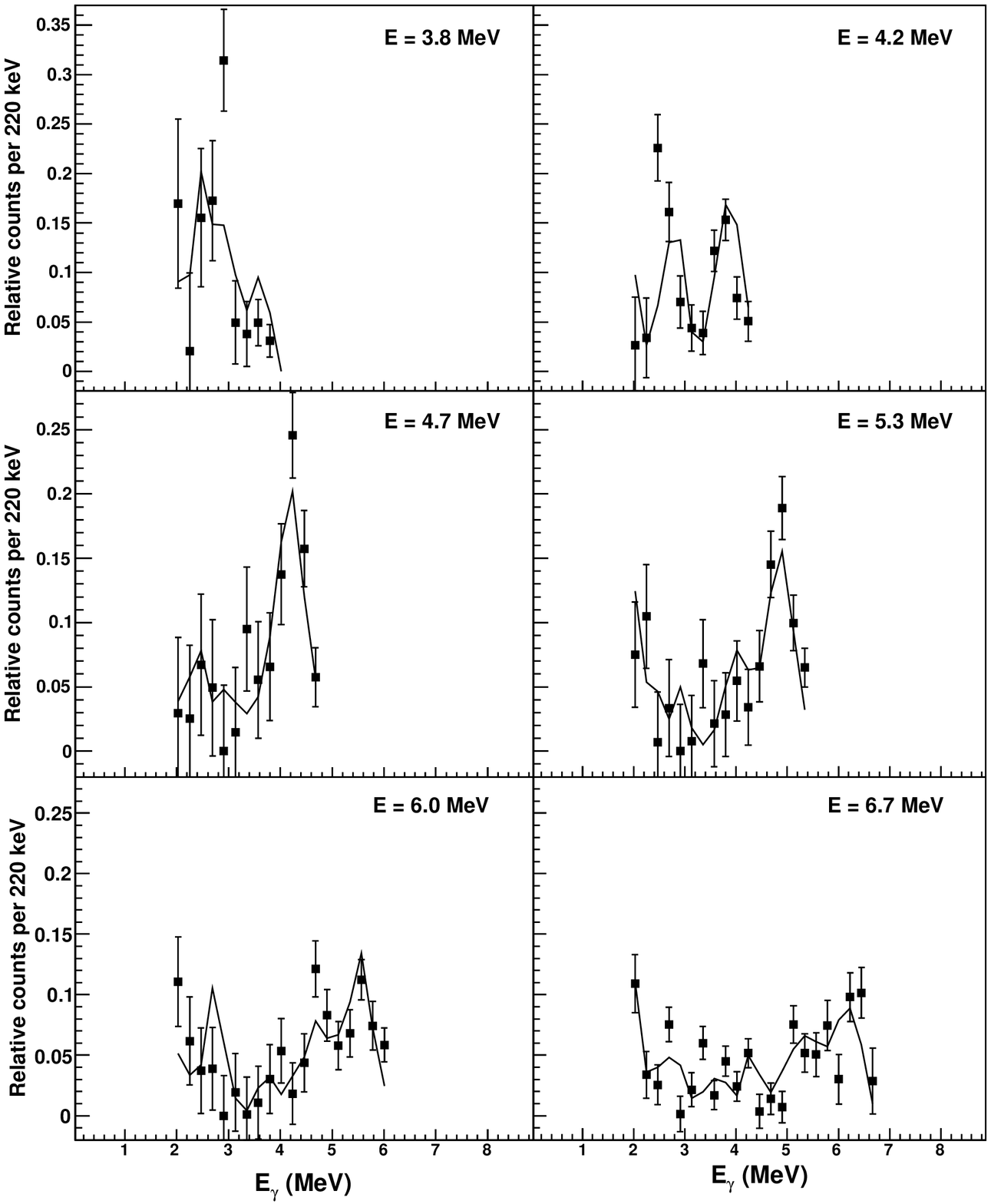}
\caption{Comparison of the normalized experimental primary $\gamma$-ray spectra for the $^{208}$Pb($^3$He,$\alpha$)$^{207}$Pb reaction (data points) at various excitation energies and the fit (solid lines) using the factorization of Eq.~(\ref{eqn:2}). The excitation energy widths are 220 keV.}
\label{fig:doesitwork}
\end {figure}

In Fig.~\ref{fig:doesitwork} such a least $\chi^2$ fit is compared with the experimental primary $\gamma$-ray matrix for the $^{208}$Pb($^3\rm{He},\alpha \gamma$)$^{207}$Pb reaction at different initial excitation energies between $E= 3.8 - 6.7$ MeV with energy bins of 220 keV. The calculated primary $\gamma$-ray spectra (solid lines) are obtained by multiplying the extracted $\rho$ and ${\cal T}$, as defined in Eq.~(\ref{eqn:3}).  

The $\gamma$-ray transmission coefficient ${\cal T}(E_\gamma)$ in Eqs.~(\ref{eqn:2}) and ~(\ref{eqn:3}) is independent of excitation energy according to the generalized Axel-Brink hypothesis~\cite{brink,axel}, stating that collective excitations built on excited states have the same properties as those built on the ground state. The average temperature in the excitation energy region studied is below 1.0 MeV for the final levels. The temperature is believed to depend on the final excitation energy by $T \propto \sqrt{E_f}$, which is a slowly varying function. Thus, we assume that using a constant temperature for the factorization of Eq.~(\ref{eqn:2}) is approximately valid in our excitation region. The error bars of the data points take into account only the statistical errors, see Ref.~\cite{schl0}. This means that neither possible shortcomings of the first-generation procedure nor a weak dependence on the excitation energy in the transmission coefficient, which means that the Axel-Brink hypothesis is not fully valid, are included in the error bars. Keeping this in mind, the comparison shown in Fig.~\ref{fig:doesitwork} indicates that the analyzing method works satisfactorily for the $^{207}$Pb nucleus.

The multiplicative functions of Eq.~(\ref{eqn:2}) give an infinite number of solutions. It has been shown in Ref.~\cite{schl0} that if one solution is known for Eq.~(\ref{eqn:2}), then the product $\rho \cdot {\cal T}$ is invariant under the transformation
\begin{equation}
\tilde{\rho}(E-E_\gamma) = A\exp[\alpha(E-E_\gamma)]\rho(E-E_\gamma),\
\label{eqn:4}
\end{equation}
\begin{equation}
\tilde{\cal T}(E_\gamma) = B\exp(\alpha \ E_\gamma) \ {\cal {T}} (E_\gamma).\
\label{eqn:5}
\end{equation}
Therefore, neither the slope nor the absolute value of ${\rho}$ and ${\cal T}$ can be determined directly from the iteration procedure. The free parameters $A$, $B$ and $\alpha$ must be determined by experimental data in order to normalize $\rho$ and $\cal T$.

\subsection{Normalization of nuclear level density}

The parameters $A$ and $\alpha$ in Eqs.~(\ref{eqn:4}) and~(\ref{eqn:5}) are obtained by fitting the inferred data points to the known discrete levels~\cite{ENSDF} at low excitation energies and to the level density at the neutron separation energy $S_n$. The level density at $S_n$ can be deduced from the Fermi-gas expression~\cite{GC} using the available proton or neutron-resonance spacing data~\cite{RIPL} and assuming that positive and negative parities contribute equally to the level density at $S_n$. For $\ell=0$ capture (s-waves), the level density $\rho_0$ becomes:
\begin{align}
\rho_0(S_n) &=   \frac{2 \sigma^2}{D_0} [ (I_t+1) \exp(-(I_t +1)^2/2\sigma^2) \nonumber  \\
&+ I_t \exp(-I_t ^2/2\sigma^2)] ^{-1}.\
\label{eqn:6.1}
\end{align}

\begin{figure}
\centering
\includegraphics[height=0.8\textheight]{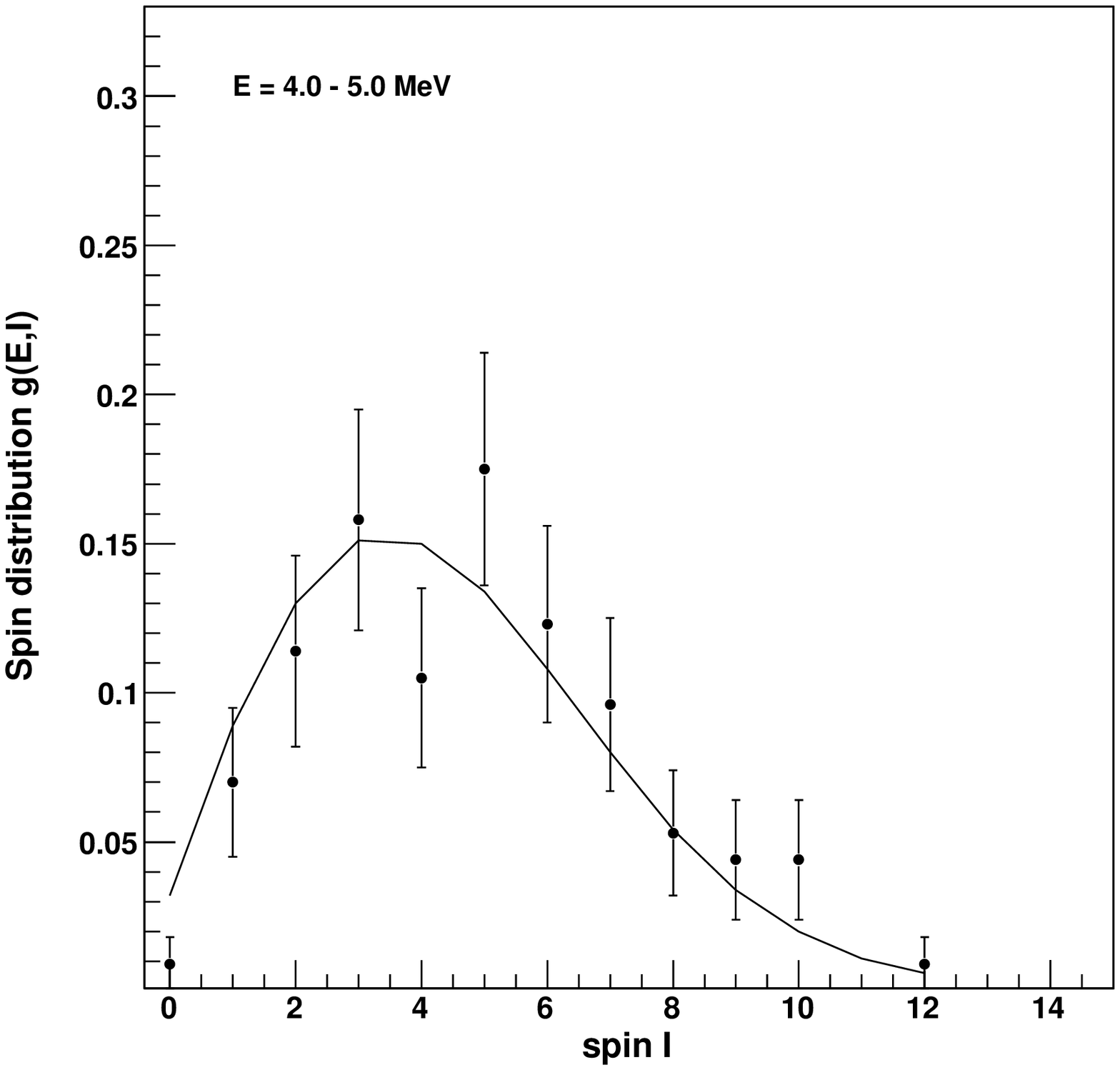}
\caption{Average experimental spin distribution (data points with error bars) of $^{206,208}$Pb compared to the spin distribution determined using Eq.~(\ref{eqn:spin}). The error bars include only statistical errors.}
\label{fig:spin}
\end {figure}

For $\ell=1$ capture (p-waves), the above equation becomes:
\begin{align}
\rho_1(S_n) &= \frac{2 \sigma^2}{D_1} [ (I_t -1) \exp(-(I_t -1)^2/2\sigma^2) \nonumber \\
 &+ I_t \exp(-I_t ^2/2\sigma^2) \nonumber \\
 &+ (I_t +1) \exp(-(I_t +1)^2/2\sigma^2) \nonumber \\
 &+ (I_t +2) \exp(-(I_t +2)^2/2\sigma^2) ] ^{-1}, \ 
\label{eqn:6.2}
\end{align}
where $D_0$ and $D_1$ are the average s- and p-wave resonance spacings. The parameter $ I_t $ is the spin of the target nucleus. For spin $ I_t =0$, the first two terms inside the bracket $[\cdots]$ of Eq.~(\ref{eqn:6.2}) should be omitted and for spin $ I_t =1/2$ and $1$ only the first term should be omitted. The spin-cut off parameter $\sigma$ is defined in Ref.~\cite{GC} by
\begin{equation}
\sigma^2 = 0.0888 A^{2/3} \sqrt{a (E-E_{\rm{pair}})}.\
\label{eqn:7}
\end{equation}
where $A$ is the mass number, $a$ is the level density parameter and $E_{pair}$ is the pairing correction parameter. The pairing correction parameter is estimated following the description of Ref.~\cite{doba}. The spin distribution of levels at one excitation energy is given by~\cite{GC}:
\begin{equation}
g(E,I) = \frac{2I+1}{2\sigma^2} \exp[-(I+1/2)^2/2\sigma^2],\
\label{eqn:spin}
\end{equation}
which is normalized to $\Sigma_I g(E,I) \sim$ 1. The spin assignments of $^{206,208}$Pb in the excitation region $E \sim$ 4 -- 5 MeV are taken from Ref.~\cite{ENSDF}. The average spin distribution of these data points are compared with the relative spin distribution determined from Eq.~(\ref{eqn:spin}), as shown in Fig.~\ref{fig:spin}. Within the statistical uncertainty of the data points the agreement of the experimental and theoretical spin distributions is very good and supports our adopted $\sigma$ in Eq.~(\ref{eqn:7}).

In Ref.~\cite{RIPL} both s-wave and p-wave resonance spacings $D_0$ and $D_1$ for the target $^{204,206,207}$Pb nuclei are given at $S_n$. The deduced level densities $\rho_0$ and $\rho_1$ at $S_n$ are listed in Table~\ref{tab:tab1}. Since the s-wave resonances in $^{207,208}$Pb are more weakly populated than the p-wave resonances, the level densities determined using $D_0$ are less reliable than those determined using $D_1$. Therefore, the $\rho_1$ level density has been used for normalizing our data.  
 
\begin{figure}
\centering
\includegraphics[height=0.8\textheight]{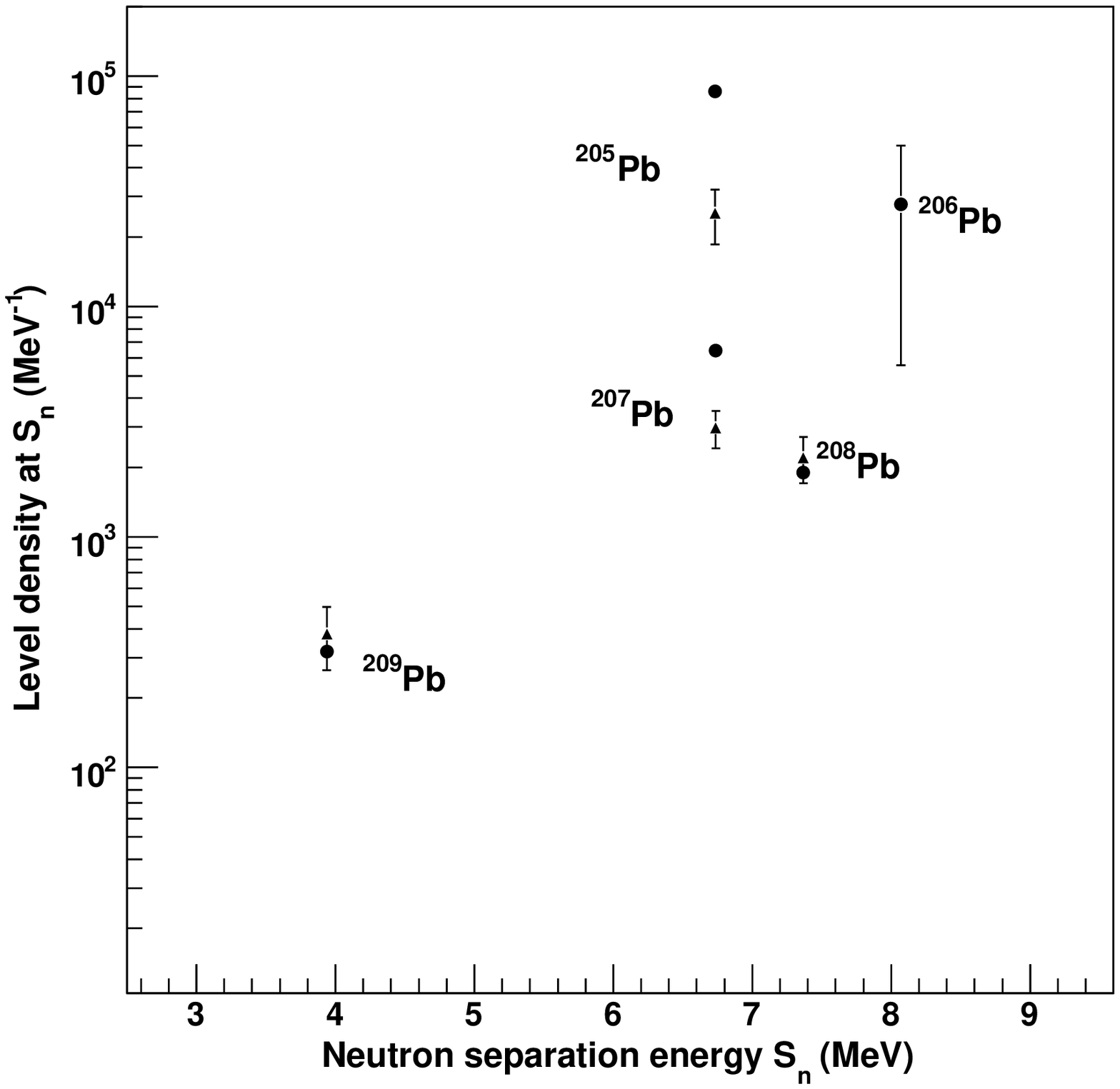}
\caption{Estimation of the level density at neutron separation energy $\rho(S_n)$ for $^{206}$Pb. The filled triangles represent the deduced $\rho(S_n)$ for $^{205,207-209}$Pb using $\ell = 1$ neutron resonance spacings~\cite{RIPL}. The filled circles are the level densities determined by Eq.~(\ref{eqn:8}) using parameters of Table~\ref{tab:tab1}.}
\label{fig:estimation}
\end {figure}

The parameters used in the analysis of $^{205-208}$Pb are summarized in Table~\ref{tab:tab1}. The level density parameter $a$ of the Pb nuclei is taken from Ref.~\cite{RIPL1} using the Gilbert and Cameron approach~\cite{GC}. Since the proton/neutron resonance spacing information for $^{206}$Pb is unavailable, we estimate the level density at the neutron separation energy using the systematics of Ref.~\cite{GC}. Figure~\ref{fig:estimation} demonstrates the estimation procedure, where the level densities of odd and even Pb nuclei are shown. In $^{208}$Pb and $^{209}$Pb the Fermi gas level densities at $S_n$ are comparable with those deduced from the neutron resonance data~\cite{RIPL}. However, in $^{205}$Pb and $^{207}$Pb the large discrepancy between the two level densities is obvious. In view of this discrepancy we have estimated an uncertainity of 80$\%$ for the adopted Fermi gas level density in $^{206}$Pb. 

The experimental level density is extracted up to $\sim$ 2 MeV below $S_n$ since the $\gamma$-rays below 2 MeV are omitted in the extraction procedure. In order to fill the gap between the data points and the level density at $S_n$ deduced from resonance data, an interpolation is required. The Fermi gas level density formula for all spins and parities
\begin {equation}
\rho_{\rm{FG}}(U) = \eta\frac{ \exp(2\sqrt{aU})}{12 \sqrt{2} a^{1/4} U^{5/4} \sigma}
\label{eqn:8}
\end{equation}
is employed for the interpolation. Here, $U = E - E_{pair}$ is the intrinsic excitation energy and $\eta$ is a constant used to adjust $\rho_{\rm{FG}}$ to the experimental level density at $S_n$ (values are given in Table~\ref{tab:tab1}). 

\begin{figure}
\centering
\includegraphics[height=0.8\textheight]{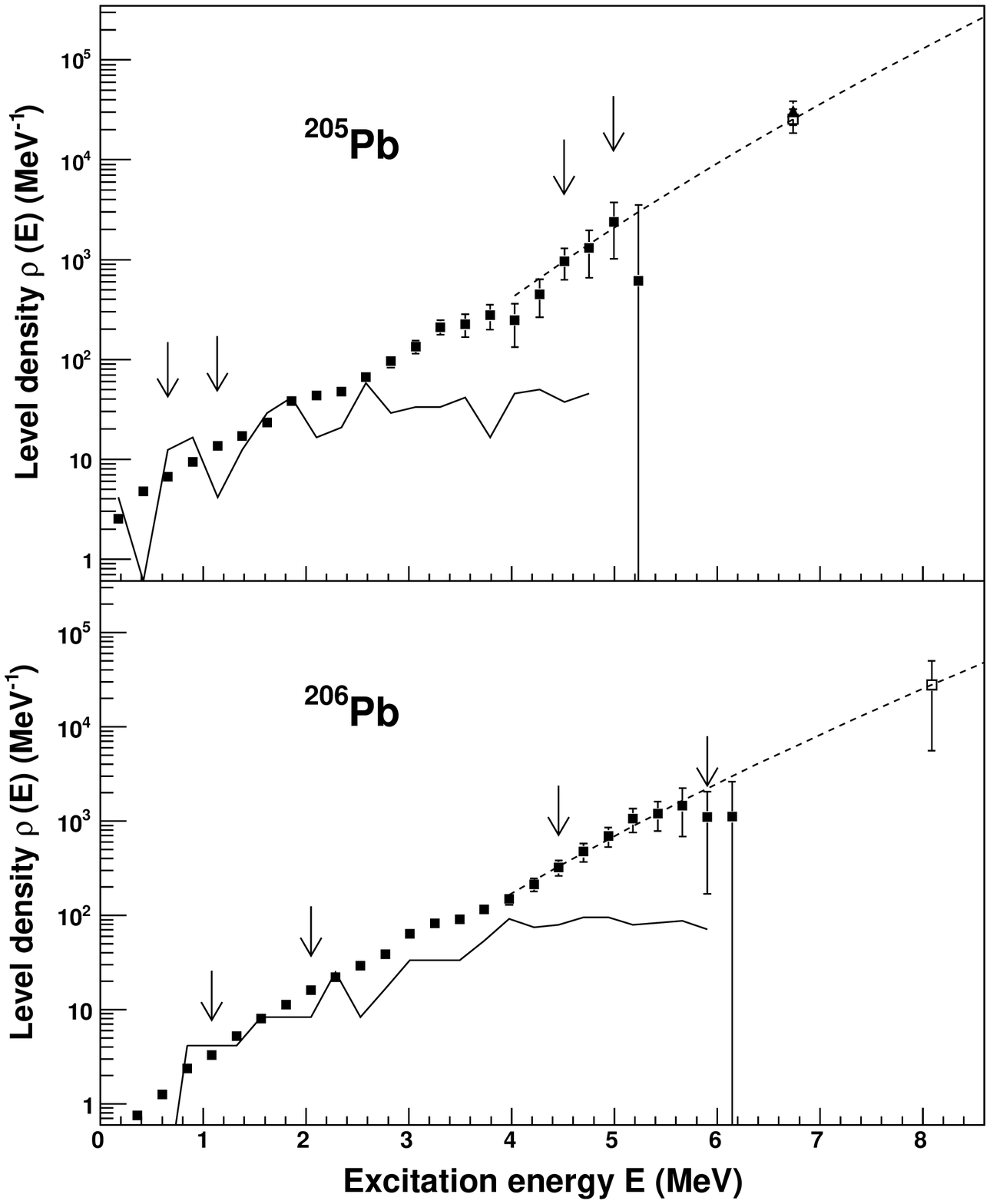}
\caption{Normalization of experimental nuclear densities (filled squares) for $^{205}$Pb and $^{206}$Pb. The arrows indicate the fitting region of data points at low excitation energies with known levels (solid line). To bridge the gap between the level density at $S_n$ (open square) and the data points, the Fermi gas level density (dashed line) is used. The filled triangle and the open square symbols at $S_n$ are based on $\ell = 0$ and $1$ neutron resonance spacing data, respectively.}
\label{fig:nld1}
\end {figure}

\begin{figure}
\centering
\includegraphics[height=0.8\textheight]{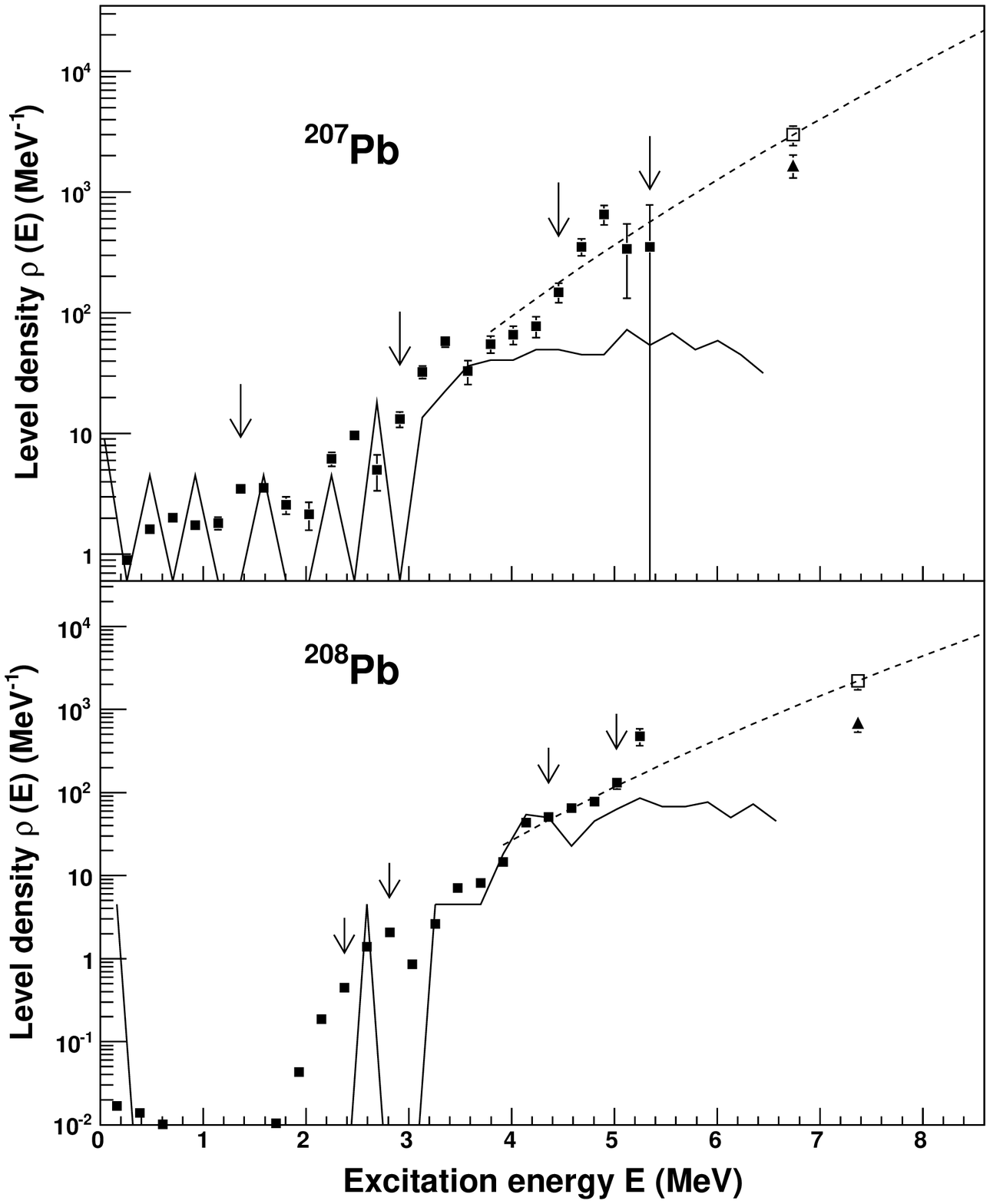}
\caption{Normalized nuclear densities (filled squares) of $^{207,208}$Pb (see caption of Fig.~\ref{fig:nld1}).}
\label{fig:nld2}
\end {figure}

Figures~\ref{fig:nld1} and~\ref{fig:nld2} show the normalized level densities extracted for $^{205,206}$Pb and $^{207,208}$Pb, respectively, up to $\sim 5 - 6$ MeV of excitation energy. The exponential increase of level density with excitation energy is evident from these figures, where the Oslo data points are shown as filled squares. The Fermi gas level densities (dashed lines) that are used for normalization at the highest energy points, are seen to describe the data points only to some extent. This feature is expected for nuclei near closed shells, having few nuclear levels. For the lead region, it is clear that the Fermi gas gives a poor description below 4--5 MeV of excitation energies.

By comparing our data with the discrete levels from spectroscopic experiments~\cite{ENSDF}, a very good resemblance is seen at lower excitation energies in all the nuclei. Local differences might be due to violation of the assumptions behind Eq.~(\ref{eqn:2}) for nuclei where level densities are low and large fluctuations of the $\gamma$-ray intensity are observed. 
In the doubly magic $^{208}$Pb the level density is low and the Oslo data agree nicely with the spectroscopic measurements~\cite{ENSDF} up to 5 MeV of excitation energy. The presence of a single unpaired neutron in $^{207}$Pb increases the level density compared to $^{208}$Pb at a given excitation energy. The Oslo data of $^{207}$Pb provide new information on level density above $E=4$ MeV. Similarly, in $^{205,206}$Pb new information on the level densities are determined at higher excitation energies where spectroscopic methods fail to find levels. 

Figure~\ref{fig:nld1} shows that the level densities of $^{205,206}$Pb are smoother functions of excitation energy with an overall higher absolute value than of $^{207,208}$Pb, where characteristic structures are prominent. This feature is expected due to the presence of two and three neutron vacancies, which provide extra degrees of freedom for the nucleons to arrange themselves for a given excitation energy.

The drop of the total level density for $E \le$ 4 MeV while going from $^{205}$Pb to $^{208}$Pb, is interpreted as a shell closure effect. The extraction of level density in $^{208}$Pb does not provide any new information above the known levels. However, these results give further confidence in the Oslo method and its applicability in this mass region, as our data show good agreement with the data found in literature. In summary, these results are gratifying and support the applicability of the Oslo method in closed-shell nuclei where the level densities are low.

\section{Thermodynamics}
The nuclear level density is closely related to the entropy $S$ of the system at a given excitation energy $E$. In fact, the level density $\rho(E)$ is directly proportional to the number of accessible levels at excitation energy $E$. In order to derive thermodynamic quantities for mesoscopic systems, it is common to use the micro-canonical or canonical ensemble. The micro-canonical ensemble is the most appropriate ensemble according to Ref.~\cite{Morissay}, since the nucleus is considered as an isolated system with well-defined energy. The micro-canonical ensemble theory will therefore be utilized for $^{205-208}$Pb, although temperature and heat capacity may become negative due to fluctuations in the entropy.\\

In the micro-canonical ensemble, the entropy $S(E)$ is identical to the partition function determined by the multiplicity of states $\Omega(E)$ at excitation energy $E$ which corresponds to the level density $\rho(E)$. The entropy in the micro-canonical ensemble can be derived as
\begin{align}
S(E) &= k_B \ln \Omega(E) \nonumber \\
	&= k_{B} \ln \frac{\rho(E)}{\rho_0}  \nonumber \\
	&= k_{B}\ln \rho(E) + S_0,\
\label{eqn:array2}
\end{align} 
where Boltzmann's constant $k_B$ is set to unity for the sake of simplicity. The value of the normalization factor $S_0$ is adjusted such that the third law of thermodynamics is fulfilled: $S \rightarrow 0$ for $T \rightarrow 0$. Since the ground-state band of the even-even $^{206}$Pb and $^{208}$Pb nuclei has $T=0$, the parameter $S_0$ used for $^{205-208}$Pb is 0.34. From the entropy, one can derive the temperature by
\begin{equation}
\frac{1}{T(E)} = \frac{\partial S}{\partial E}. \
\label{eqn:array3} 
\end{equation} 

\begin{figure}
\centering
\includegraphics[height=15cm]{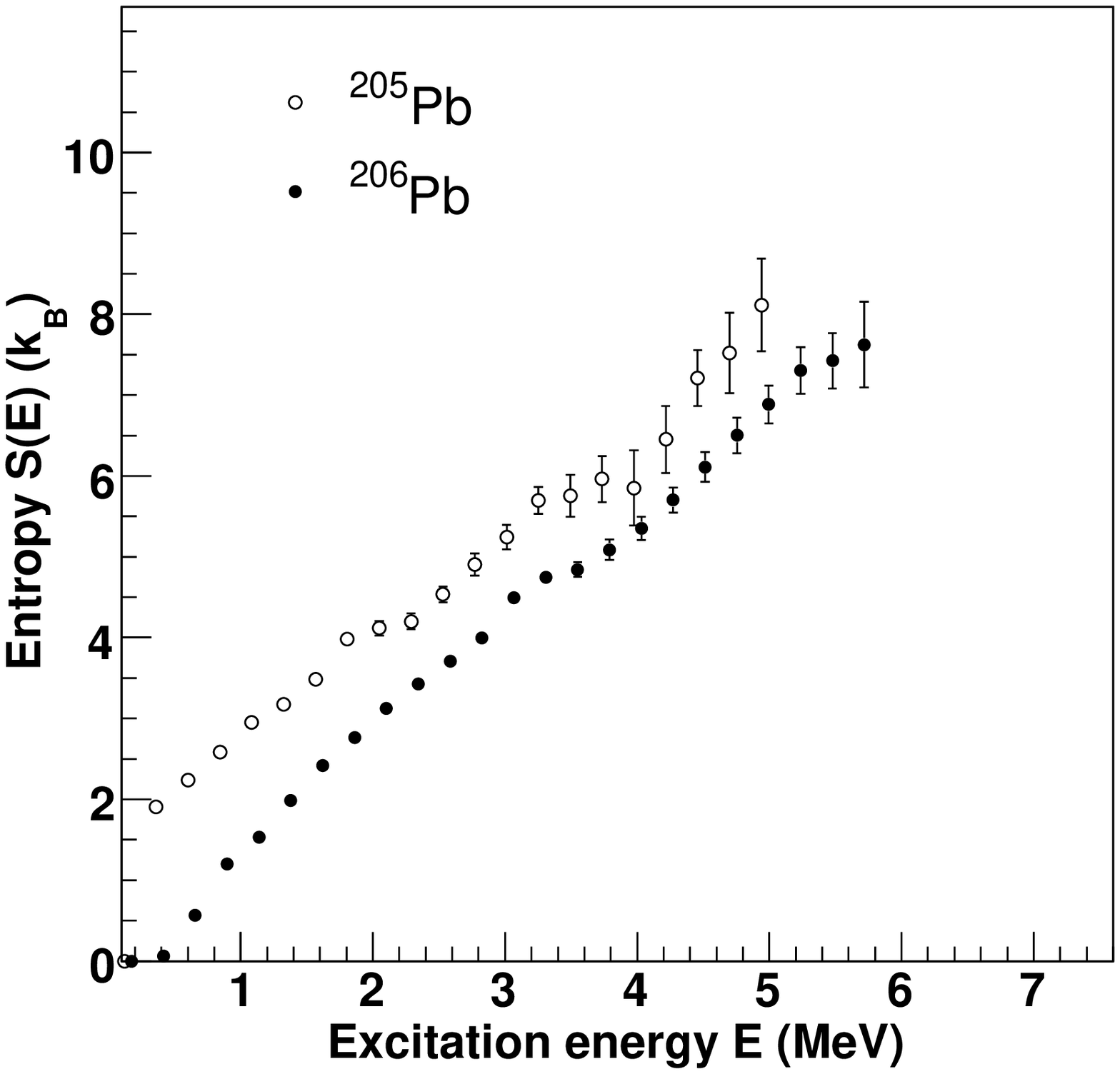}
\caption{Micro-canonical entropy for $^{205,206}$Pb.}
\label{fig:entropy2}
\end {figure}

\begin{figure}
\centering
\includegraphics[height=15cm]{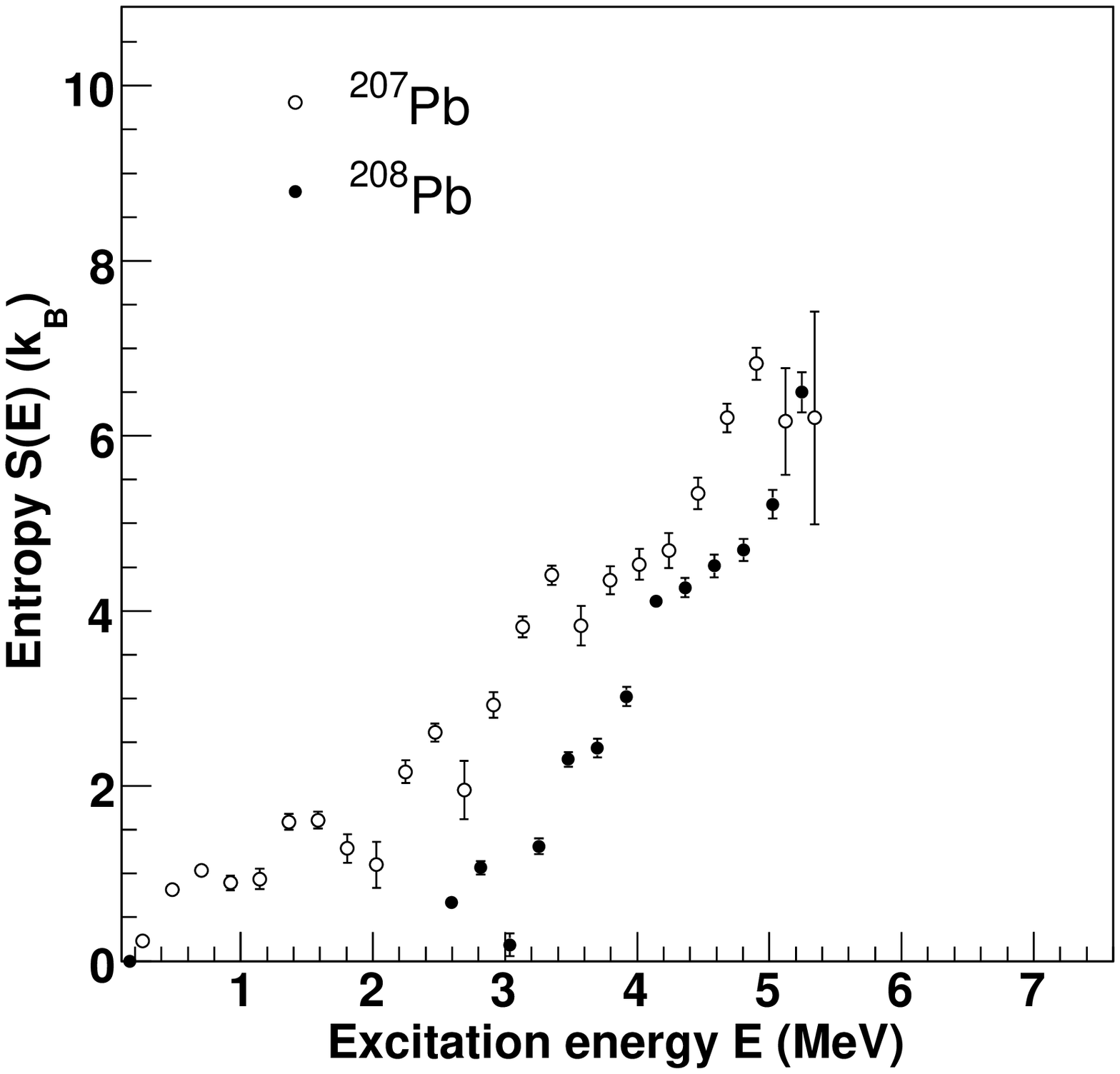}
\caption{Micro-canonical entropy for $^{207,208}$Pb.}
\label{fig:entropy1}
\end {figure}

\begin{figure}
\centering
\includegraphics[height=15cm]{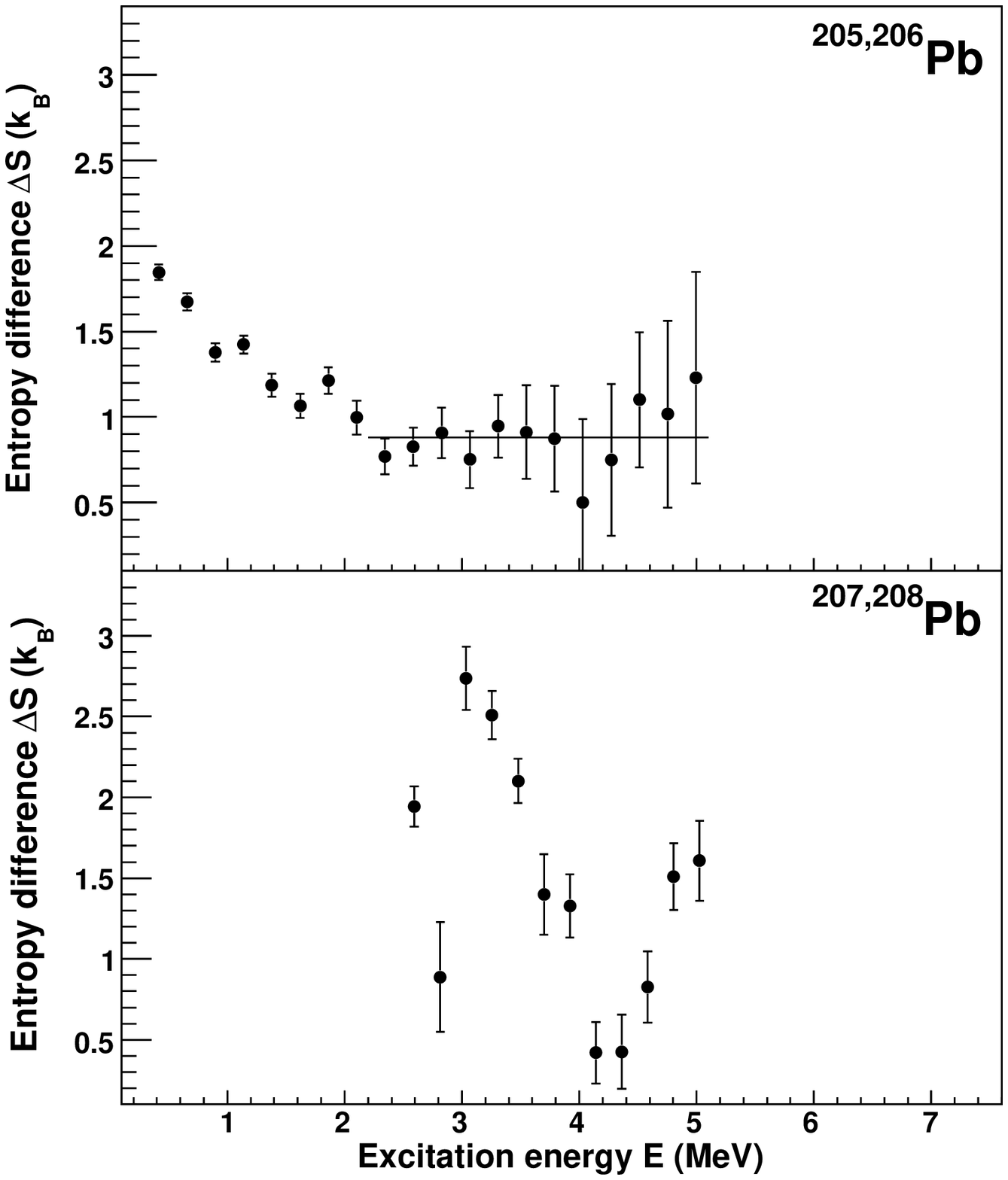}
\caption{Difference in entropy between $^{205}$Pb and $^{206}$Pb (upper panel) and between $^{207}$Pb and $^{208}$Pb (lower panel). The entropy difference is given by $\Delta S = S_{\mathrm{eo}} - S_{\mathrm{ee}}$, where $S_{\mathrm{eo}}$ and $S_{\mathrm{ee}}$ is the entropy of the even-odd and even-even nucleus, respectively.}
\label{fig:delS2}
\end {figure}

In Figs.~\ref{fig:entropy2} and~\ref{fig:entropy1} the micro-canonical entropies of $^{205,206}$Pb and $^{207,208}$Pb are shown, respectively. The entropies of $^{207, 208}$Pb are seen to vary strongly with excitation energy. For $^{208}$Pb, the first vibrational 3$^{-}$ state appears at 2.6 MeV, before the two quasi-particle regime which is seen to enter at $E>$ 3.5 MeV. Figure~\ref{fig:delS2} shows that the difference in entropy between $^{207}$Pb and $^{208}$Pb, $\Delta S = S($$^{207}$Pb$)-S($$^{208}$Pb$)$, is fluctuating between $0.3 - 2.0$ in the excitation energy region of $E=3.0 - 5.7$ MeV. This strong variation in entropy is due to the few available single-particle orbitals for the unpaired neutrons below the closed shell. The strong fluctuation of entropy in $^{207,208}$Pb with excitation energy makes it difficult to determine other thermodynamic quantities such as temperature.

In Fig.~\ref{fig:delS2} the average difference in entropy between $^{205}$Pb and $^{206}$Pb is shown to lie around 0.9 for the excitation energy region between $E = 2 - 5$ MeV. The entropy difference of $^{205}$Pb and $^{206}$Pb is less fluctuating than that observed between $^{207}$Pb and $^{208}$Pb, thus indicating the departure of shell closure effects in $^{205,206}$Pb. The entropy difference between $^{205}$Pb and $^{206}$Pb is about one half of the value observed in rare earth nuclei ($\Delta S \sim$ 2)~\cite{bagheri}.  

\begin{figure}
\centering
\includegraphics[height=15cm]{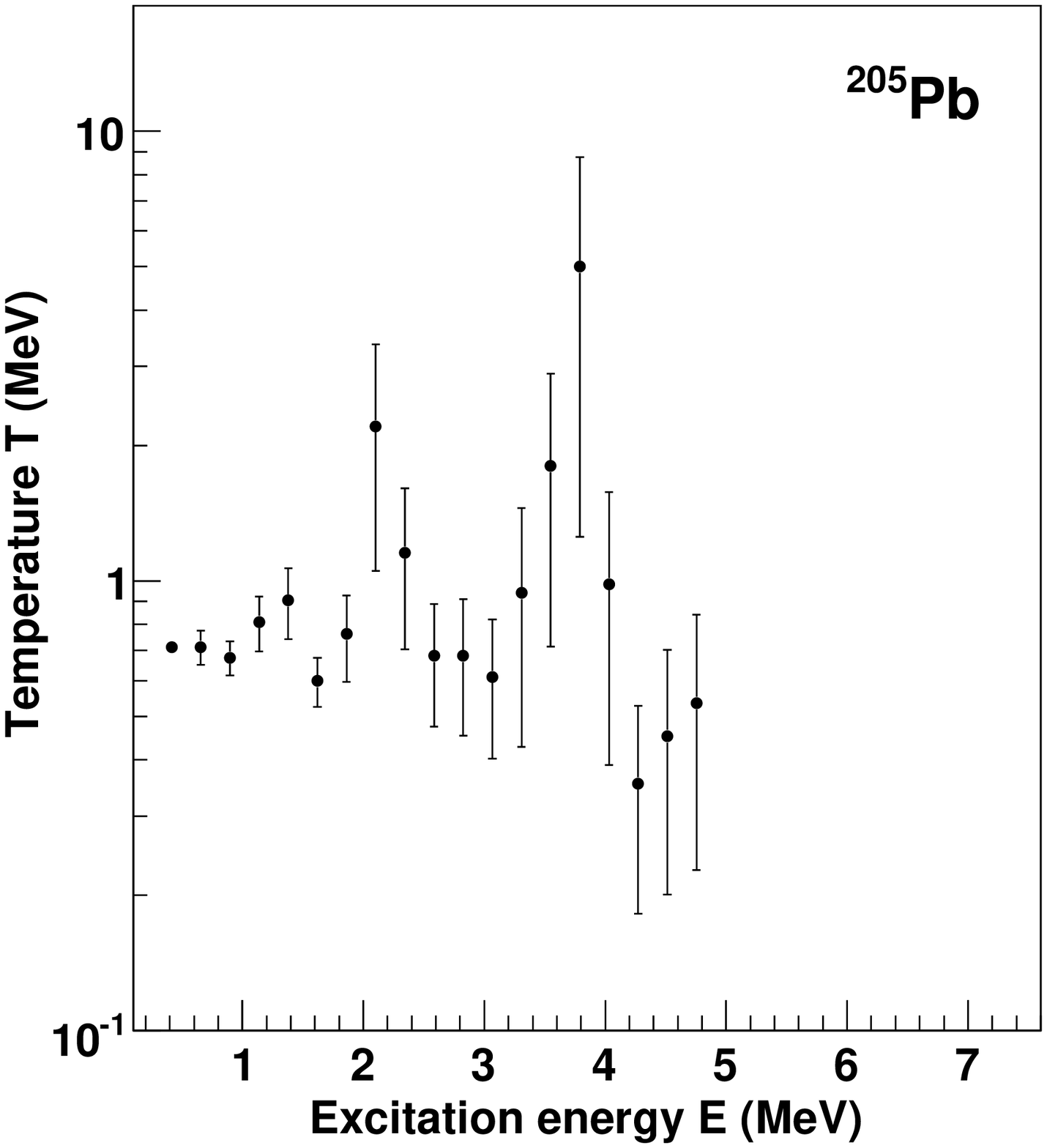}
\caption{Temperature as a function of excitation energy in $^{205}$Pb.}
\label{fig:temp1}
\end {figure}

\begin{figure}
\centering
\includegraphics[height=15cm]{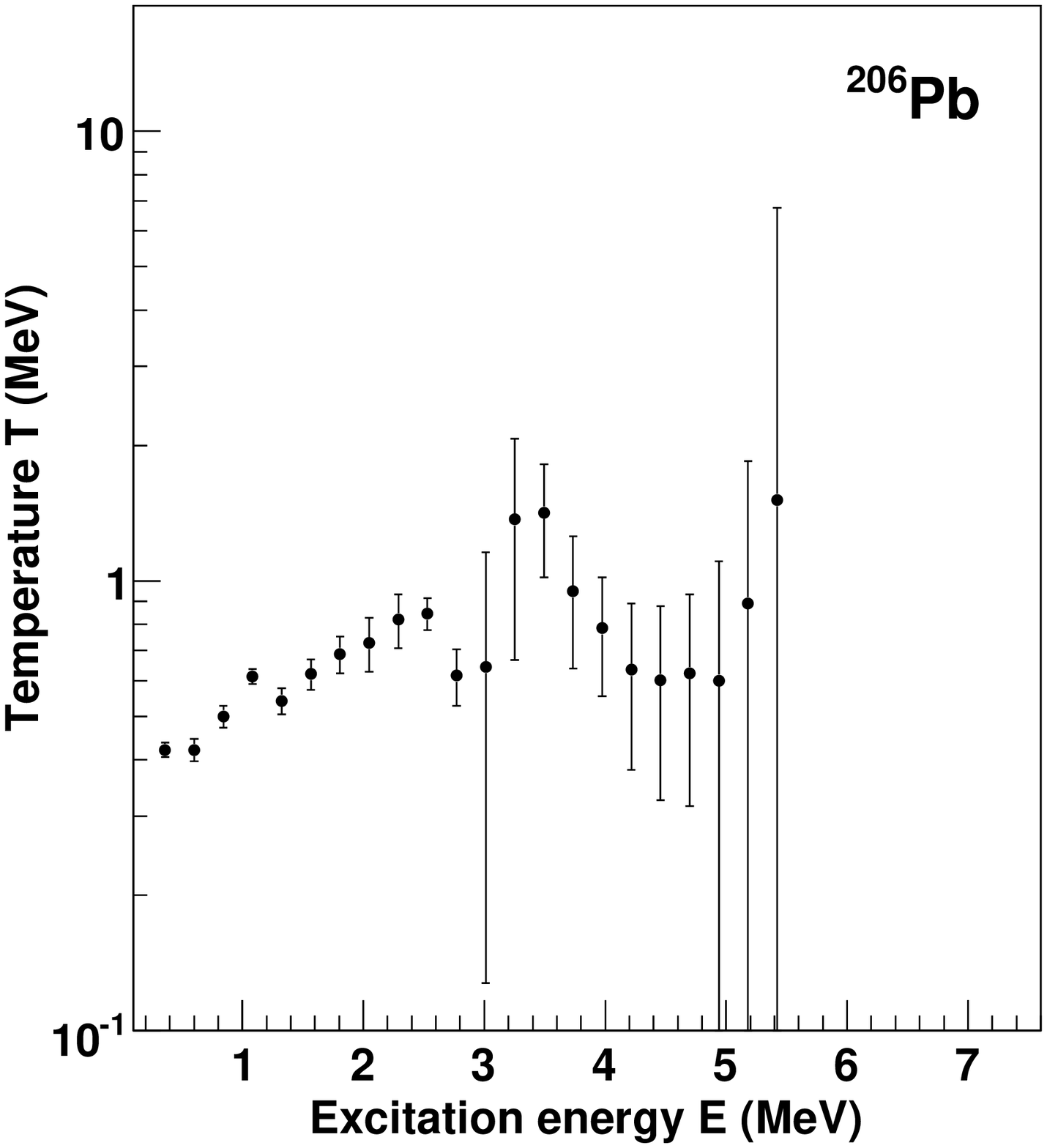}
\caption{Temperature as a function of excitation energy in $^{206}$Pb.}
\label{fig:temp2}
\end {figure}

The micro-canonical temperature can be extracted from the entropy as given by Eq.~(\ref{eqn:array3}). Since the level densities in $^{207,208}$Pb are low and show large fluctuations, the temperature extraction for these nuclei is probably not reliable. The micro-canonical temperatures of $^{205}$Pb and $^{206}$Pb are shown in Figs.~\ref{fig:temp1} and ~\ref{fig:temp2}, respectively. For $^{206}$Pb, the temperature extraction gives rise to large bump structures. These structures in the temperature arise from the differentiation of the entropy (see Eq.~(\ref{eqn:array3})), and can be interpreted as the breaking of nucleon pairs. When nucleon pairs are broken, new degrees of freedom open up leading to an increase of $\rho(E)$ and decrease in the temperature $T(E)$.

For $^{205}$Pb, the situation is not so clear since this nucleus has an unpaired neutron which gives a smoother entropy and thus less structures in the temperature. Bumps indicating the pair-breaking process might also be seen here, but one can also fit a constant temperature to the data points in Fig.~\ref{fig:temp1}. By doing so, we get an average temperature of 0.9(1) MeV for $0.4 \leq E \leq 5.0$ MeV, which is in agreement with an average temperature of $T=0.81(4)$ MeV~\cite{E&B}. 

In $^{205}$Pb and $^{206}$Pb one can observe the first drop in temperature at $E \approx$ 2.0 MeV and $E \approx$ 2.5 MeV, respectively. These excitation energies can be compared with the energy amount necessary to break a neutron Cooper pair, namely 2$\Delta_n$, where $\Delta_n$ is the neutron pair gap parameter. The description of Ref.~\cite{GC} gives 2$\Delta_n$ = 1.8 MeV for $^{205,206}$Pb. The observed locations of pair-breaking lie at energies somewhat higher than 2$\Delta_n$. This difference could be due to the large energy spacing between the single-particle orbitals. However, the second peak in $^{205}$Pb and $^{206}$Pb is observed at $E \approx$ 3.8 and 3.7 MeV, respectively. These peaks may be due to proton-pair breaking, neutron-pair breaking, or because the proton (or neutron) passes the energy gap between major shell gaps.

\section{Gamma-ray strength functions}

\begin{figure}
\includegraphics[height=10cm]{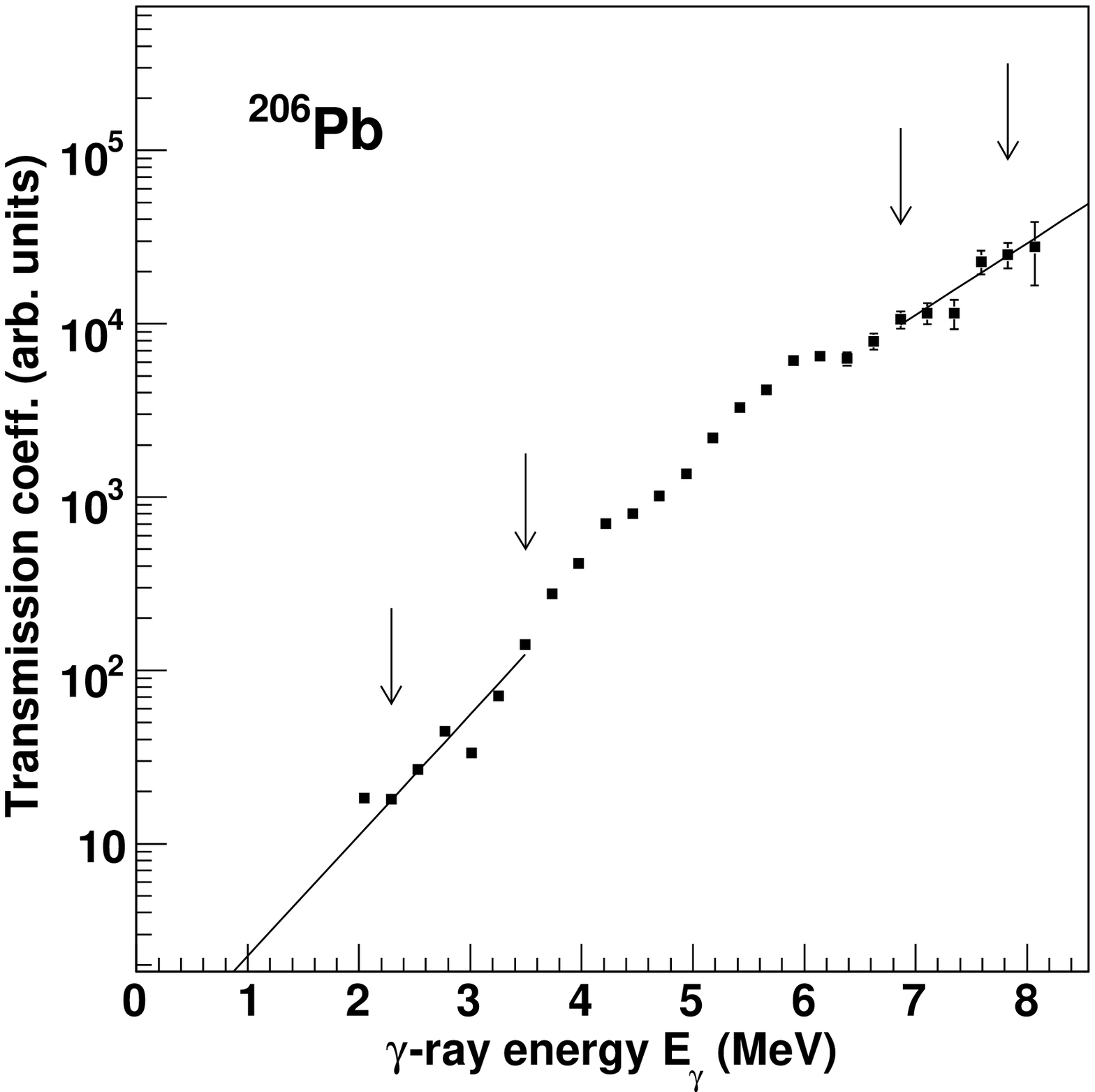}
\caption{Transmission coefficient $\cal {T}$ in arbitrary units for $^{206}$Pb. An exponential function (solid line) is fitted to the data points between the arrows at low and high $\gamma$-ray energies.}
\label{fig:tau}
\end {figure}

The $\gamma$-ray transmission coefficient $\cal {T}$ in Eq.~(\ref{eqn:2}) is connected to the electromagnetic decay properties of the nucleus and is expressed as the sum of all  the $\gamma$-ray strength functions $f_{XL}$ for transitions with electromagnetic character $X$ and multipolarity $L$, given by:
\begin{equation}
{\cal T}(E_\gamma) = 2\pi\sum_{XL} E_\gamma^{2L+1}f_{XL}(E_\gamma),\
\label{eqn:9}
\end{equation}
where $E_\gamma$ is the transition energy, and the slope correction $\exp(\alpha E_\gamma)$ from Eq.~(\ref{eqn:4}) has already been included in $\cal T$. In Fig.~\ref{fig:tau}, $\cal {T}$ for $^{206}$Pb is shown, where the absolute normalization $B$ remains to be determined. This normalization will be discussed in the following.

As already mentioned in sect.~III, the methodological difficulties in the primary $\gamma$-ray extraction prevent the determination of $\rho$ for $E > S_n-2$ MeV and ${\cal T}$ for $E_\gamma < 2$ MeV. The level densities were extrapolated with the Fermi gas level density using Eq.~(\ref{eqn:8}). However, the transmission coefficients are extrapolated using an exponential form, as shown in Fig.~\ref{fig:tau}. 

It is assumed that the main contributions to the function ${\cal T}(E_\gamma)$ are $E1$ and $M1$ $\gamma$-ray transitions in the statistical region. Further, if one assumes that the number of accessible levels of positive and negative parity are equal for any energy and spin, one finds
\begin{equation}
\rho(E-E_\gamma, I_f, \pm\pi_f) = \frac {1}{2} \rho(E-E_\gamma, I_f). \
\label {eqn:10}
\end{equation}
Clearly, this assumption does not hold true for all excitation energy regions in the Pb nuclei considered here. However, by applying these assumptions, the experimental $\gamma$-ray transmission coefficient of Eq.~(\ref{eqn:9}) yields
\begin{equation}
\ B {\cal T}(E_\gamma) = 2 \pi [f_{E1}(E_\gamma)+f_{M1}(E_\gamma)] E_{\gamma}^3,\
\label{eqn:11}
\end{equation}
where $B$ is an unknown factor that gives the absolute normalization of the $\gamma$-ray strength function. We might expect a potential error in the absolute normalization of the $\gamma$-ray strength functions in the Pb region due to equal-parity assumption used above. 

The average total radiative width $\langle\Gamma_\gamma\rangle$ of levels with energy $E$, spin $I$ and parity $\pi$ is given by~\cite{kopecky}: 
\begin{align}
\langle\Gamma_\gamma(E, I, \pi)\rangle&=\frac{1}{2\pi\rho(E, I, \pi)} \sum_ {XL}\sum_{I_f, \pi_f}\int_0^{E}{\mathrm{d}}E_\gamma \mathcal{T}_{XL}(E_\gamma)\nonumber\\
& \times \rho(E - E_\gamma, I_f, \pi_f),
\label{eqn:16B}
\end{align}
where the summation goes over spins of final levels $I_f$ with parities $\pi_f$. Using the assumptions discussed above, one can determine the average total radiative width $\langle\Gamma_\gamma\rangle$ for neutron resonances. For $\ell=0$ capture (s-waves), the populated spins are $I = |I_t \pm 1/2|$, where $I_t$ represents the spin of the target nucleus in the $(n,\gamma)$ reaction. For $\ell=1$ capture (p-waves), the populated spins are $I = |I_t \pm 1/2 \pm 1 |$. The parity is determined by the target parity $\pi _t$ and $\ell$ by $\pi =\pi _t (-1)^{\ell}$. The average total radiative width is described at $E = S_n$ by
\begin{align}
\langle\Gamma_\gamma(S_n, I)\rangle = &\frac{1}{4\pi \rho(S_n, I, \pi)} \int_0^{S_n}{\mathrm{d}}E_{\gamma} \nonumber\\
\times &B \mathcal{T}(E_\gamma)\rho(S_n - E_\gamma) \nonumber\\
\times &\sum_ {J=-1}^{1}{g(S_n-E_\gamma, I + J)}.
\label{eqn:17}
\end{align}
The factor $B$ can thus be determined from the average total radiative width $\langle\Gamma_\gamma\rangle$ of the compound states. Ref.~\cite{RIPL} provides the experimental $\langle\Gamma_\gamma\rangle$ for $^{205}$Pb, while the $\langle\Gamma_\gamma\rangle$ in $^{207}$Pb is taken from Ref.~\cite{mughabgab}. 

In order to normalize the $\gamma$-ray strength function of $^{208}$Pb the above mentioned method is difficult to apply since the average total radiative width at $S_n$ cannot be found in the literature. However, there are discrete neutron resonance data~\cite{MCC} for $^{208}$Pb on the $E1$ and $M1$ $\gamma$-ray strength functions at $E_\gamma$ = 7.5 MeV. The Oslo data for $^{208}$Pb is thus normalized with the data of Ref.~\cite{MCC}. In addition, the ($\gamma$,n) photonuclear cross-section data of Ref.~\cite{yadfys} has been used to confirm the normalization of the $\gamma$-ray strength function of $^{208}$Pb. The following relation~\cite{RIPL} can be used to deduce the strength function from the cross-section, asumming that dipole radiation is dominant:
\begin{equation}
f(E_\gamma)= \frac {1} {3\pi^2 \hbar^2 c^2} \frac {\sigma(E_\gamma)}{E_\gamma}. \
\label{eq:14}
\end{equation}

\begin{figure}
\includegraphics[height=15cm]{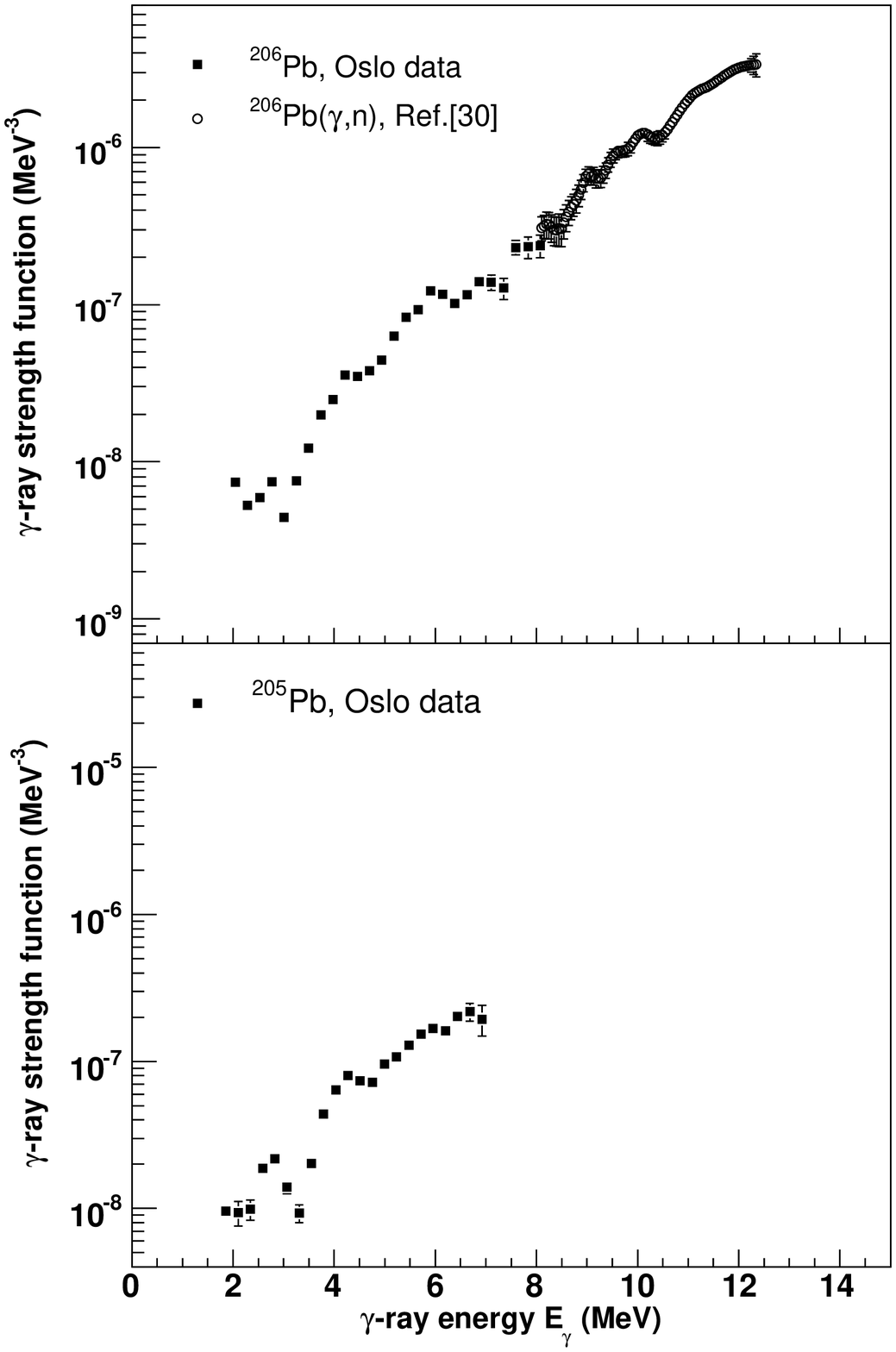}
\caption{Normalized $\gamma$-ray strength functions for $^{205,206}$Pb as a function of $\gamma$-ray energy. The Oslo data (filled squares) of $^{206}$Pb are normalized to photonuclear data (open circles) taken from Ref.~\cite{yadfys}.}
\label{fig:rsf2}
\end {figure}

\begin{figure}

\includegraphics[height=15cm]{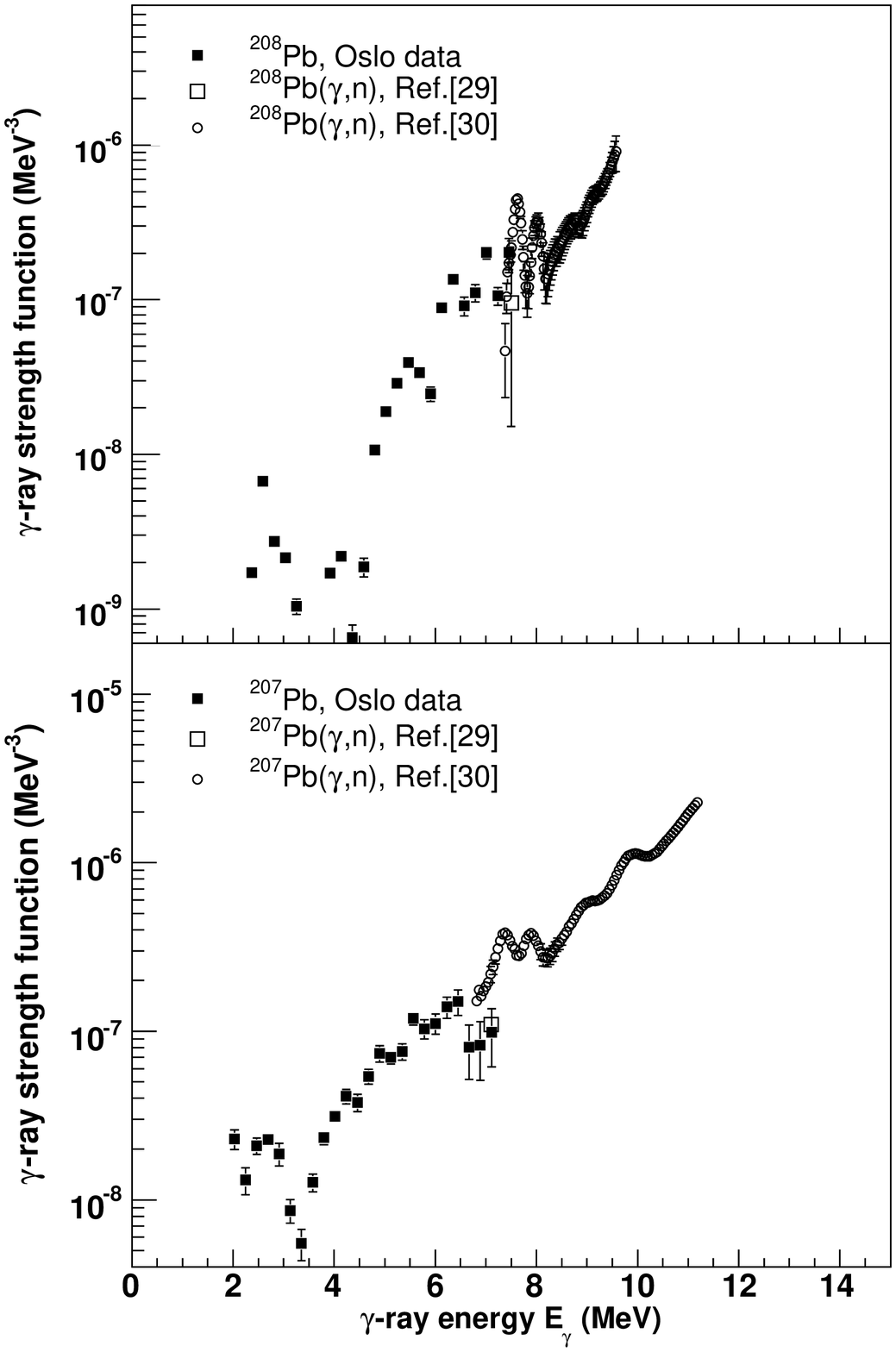}
\caption{Normalized $\gamma$-ray strength functions for $^{207,208}$Pb. The Oslo data (filled squares) of $^{207}$Pb is normalized using the data of Ref.~\cite{mughabgab}. The $^{208}$Pb nucleus is normalized by scaling the Oslo data with the sum of $f_{E1}+f_{M1}$ (open square) taken from Ref.~\cite{MCC} and the photonuclear data (open circles) of Ref.~\cite{yadfys}.}
\label{fig:rsf1}
\end {figure}

Since $^{205}$Pb is unstable, the resonance spacings and $\langle\Gamma_\gamma\rangle$ have not been measured for $^{206}$Pb. Therefore, we utilized the ($\gamma$,n) cross-section data~\cite{yadfys} to normalize the $\gamma$-ray strength function of $^{206}$Pb. The Oslo data are scaled to match the data of Ref.~\cite{yadfys}, giving the absolute normalization of the $\gamma$-ray strength function of $^{206}$Pb. The normalized $\gamma$-ray strength functions of $^{205,206}$Pb and $^{207,208}$Pb are shown in Figs.~\ref{fig:rsf2} and~\ref{fig:rsf1}, respectively.

For $^{207}$Pb, the resonance data~\cite{MCC} for electric and magnetic dipole radiation at $E_\gamma$ = 7.1 MeV have been used to make a consistency check for the normalized $\gamma$-ray strength function. Moreover, the ($\gamma$,n) data of Ref.~\cite{yadfys} for $^{207}$Pb at $E_\gamma > S_n$ are also drawn in Fig.~\ref{fig:rsf1} for comparison. The agreement of these strength functions with our extracted data is gratifying, and indicates that the normalization procedure works well in this case, in spite of the questionable assumption of equal parity distribution described in Eq.~(\ref{eqn:10}).

The ($\gamma$,n) data for $^{206-208}$Pb in Ref.~\cite{yadfys} display bumps and structures that increase in magnitude when approaching $S_n$. Similar structures are present in our data below $S_n$. These intermediate structures are observed in the Oslo data at $\gamma$-ray energies 5.6, 6.3 and 7.1 MeV in $^{208}$Pb. Similarly, in $^{207}$Pb these intermediate structures have been seen at 4.3, 5.0, 5.6 and 6.3 MeV. For $^{205}$Pb and $^{206}$Pb these structures are observed at $\gamma$-ray energies of 4.3, 6.0 and 6.3 MeV. It is hard to interpret these structures without knowing the electromagnetic character and the multipolarity of transitions in this region.  However, it is likely that the bumps are due to enhanced single-particle transitions at certain $\gamma$-ray energies, reflecting the shell structure below $S_n$. The Oslo method does not provide any information on the multipolarity of these structures, so to investigate it further, it would be necessary to perform complementary experiments such as (n,$\gamma\gamma$) measurements analyzed with the two-step cascade method~\cite{2step}.

The above extracted $\gamma$-ray strength functions depend on the normalization procedure chosen. The slope of the strength function is sensitive to the resonance data at $S_n$, which have been taken from the literature. In $^{206}$Pb the uncertainty of this data play a central role in the normalization of our data. The adopted values influence both the slope of the level density as well as the slope of the strength function. In $^{206}$Pb, the uncertainty in the value of $\alpha$ (see Eq.~(\ref{eqn:9})) could be more than 80$\%$ due to this uncertainty. In addition, the absolute strength ($B$ of Eq.~(\ref{eqn:11})) is uncertain by a factor of 2 -- 3. In the case of $^{208}$Pb, the oscillating shape of the strength function makes the absolute normalization between our data and ($\gamma$,n) data~\cite{MCC,yadfys} very difficult, as seen in Fig.~\ref{fig:rsf1}.

\section{Models for E1 and M1 Transitions}
A number of models describing the electric dipole $\gamma$-ray strength function have been described in~\cite{RIPL}. The simplest of these is the Standard Lorentzian model (SLO), which describes the $E1$ strength as a Lorentzian shape. The model is temperature independent and is given by:
\begin{equation}
f_{\mathrm{SLO}}(E_\gamma) = \frac {1} {3\pi^2 \hbar^2 c^2} \sigma_{E1} \Gamma_{E1} \frac{E_\gamma\Gamma_{E1}}{(E_\gamma^2-E_{E1}^2)^2+(\Gamma_{E1} E_\gamma)^2}, \
\label{eq:15}
\end{equation}
where the Lorentzian parameters $\sigma_{E1}$, $E_{E1}$ and $\Gamma_{E1}$ are the peak cross section, energy and width of the giant dipole resonance, respectively, and are usually derived from photonuclear experiments. However, it has been shown ~\cite{lone, MCC, kopecky, becvar, coceva} that the SLO model overestimates the photonuclear data away from the giant dipole resonance (GDR) centroid in many nuclei.

One model that incorporates the temperature dependency of the $\gamma$-ray dipole strength is the Enhanced Generalized Lorentzian model (EGLO)~\cite{RIPL}. Considering spherical nuclei, EGLO is defined as
\begin{align}
f_{\mathrm{EGLO}}(E_\gamma)  = & \frac {1} {3\pi^2 \hbar^2 c^2} \sigma_{E1} \Gamma_{E1}  \nonumber\\
 & \bigg\{ \frac{E_\gamma \Gamma_k(E_\gamma, T)}{(E_\gamma^2-E_{E1}^2)^2+(E_\gamma \Gamma_k(E_\gamma, T))^2} \nonumber\\
 & +0.7 \frac{\Gamma_k(E_\gamma=0, T)}{E_{E1}^3} \bigg\}, \
\label{eq:19}
\end{align}
where $T$ is the temperature of the final states determined by $T = \sqrt{(U/a)}$. The energy and temperature dependent width $\Gamma_k(E_\gamma, T)$ is defined analytically as
\begin{equation}
\Gamma_k(E_\gamma, T) = K(E_\gamma) \frac{\Gamma_{E1}}{E_\gamma^2}[E_\gamma^2 + (2\pi T)^2],\\
\label{eqn:16}
\end{equation}
where
\begin{equation}
K(E_\gamma) = \kappa + (1-\kappa) \frac{E_\gamma - E_0}{E_{E1} - E_0}. \\
\label{eqn:16A}
\end{equation}
is an empirical function. Here we use $E_0 = 4.5$ MeV and the enhancement factor $\kappa$, given by~\cite{kappa}, 
\begin{equation}
\kappa=
\begin{cases}
 1  & \text{if $A<148$},\\
1+0.09(A-148)^2 \exp(-0.18(A-148)) & \text{if $A \ge 148$},\\
\end{cases}
\end{equation}
for the Fermi gas model. These expressions are developed in the framework of the collisional damping model for $E_\gamma < E_{E1}$ and hold for $T <$ 2 MeV. 

The magnetic dipole $M1$ radiation is described by a Lorentzian based on the existence of a giant magnetic dipole resonance (GMDR)~\cite{GMDR},
\begin{equation}
f_{M1}(E_\gamma) = \frac{1}{3\pi^2 h^2 c^2} \frac{\sigma_{M1}E_\gamma \Gamma_{M1}^2}{(E_\gamma^2 - E_{M1}^2)^2 + E_\gamma^2 \Gamma_{M1}^2}.\
\label{eq:22}
\end{equation}
where $\sigma_{M1}$, $\Gamma_{M1}$, and $E_{M1}$ are the GMDR parameters deduced from the systematics given in~\cite{RIPL}.\\

\begin{figure}

\includegraphics[height=15cm]{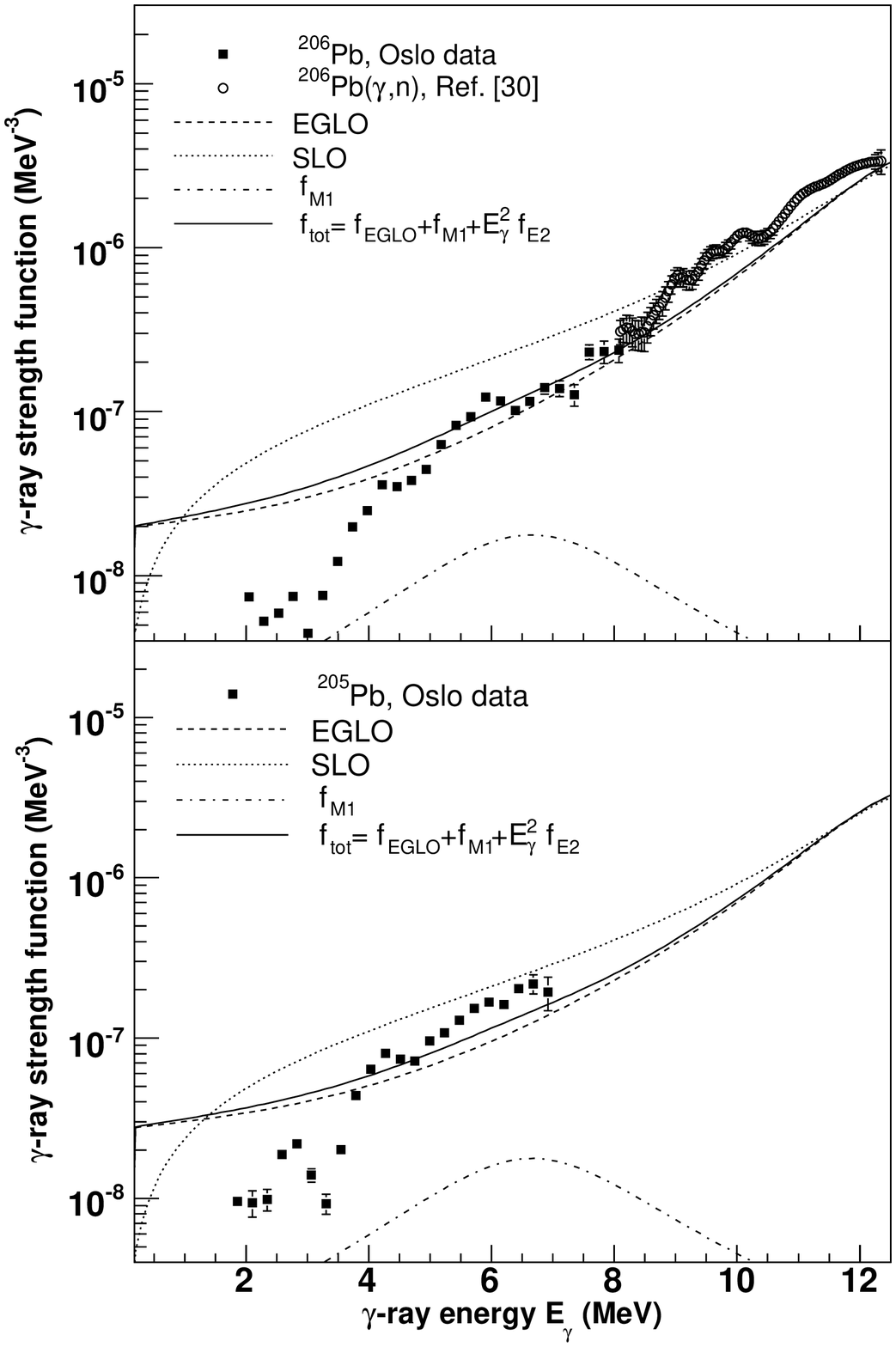}
\caption{Normalized $\gamma$-ray strength functions for $^{206}$Pb and $^{205}$Pb. The Oslo data (filled squares) are compared with the SLO (dotted line) and EGLO (dashed line) model predictions. The $M1$ strength function is shown by the dashed-dotted line. The solid line is the total strength function which is the sum of the EGLO model, $E2$ and $M1$. The photonuclear cross-section data from Ref.~\cite{yadfys} are shown as open circles.}
\label{fig:model2}
\end {figure}

\begin{figure}
\includegraphics[height=15cm]{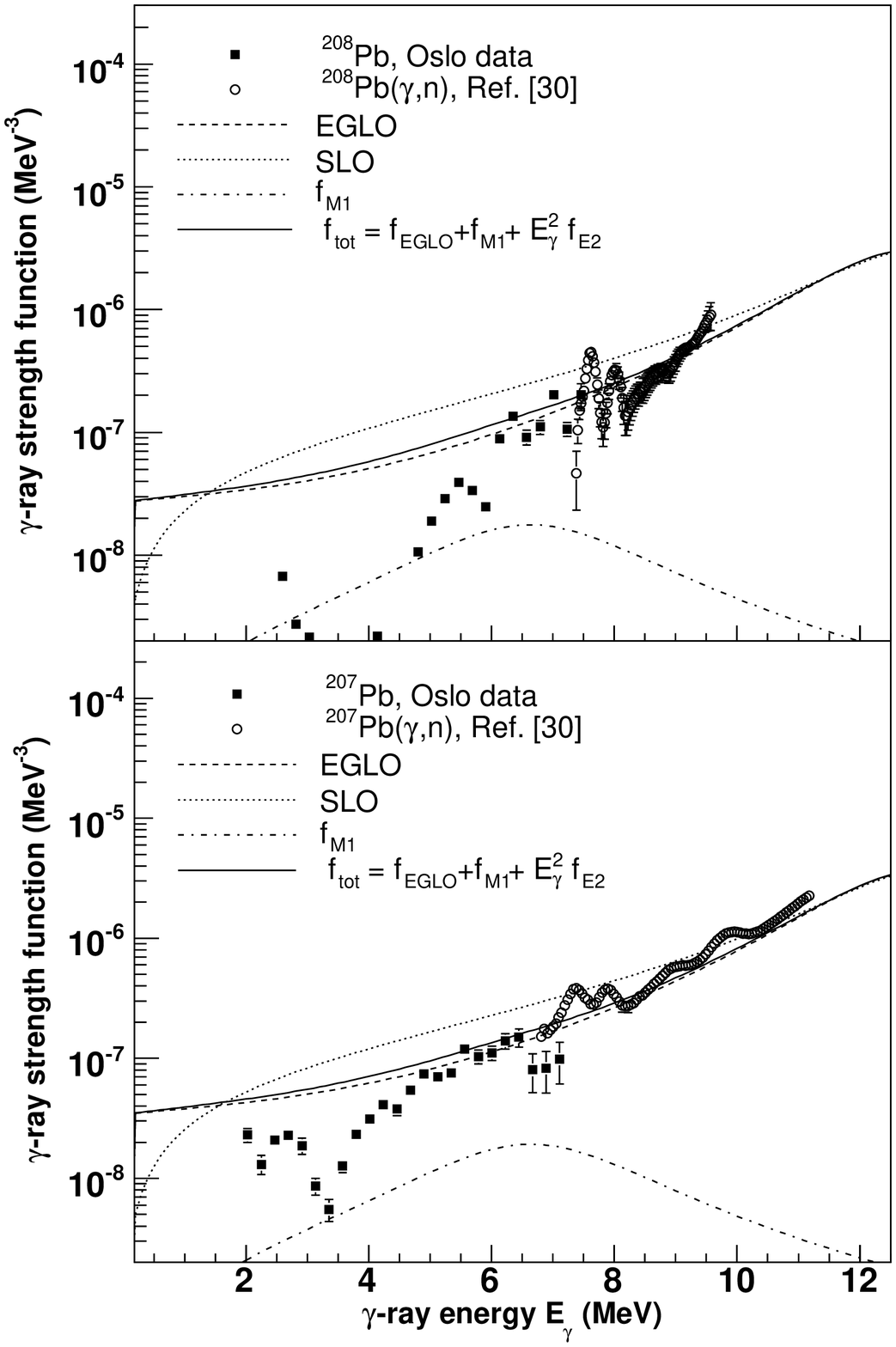}
\caption{Normalized $\gamma$-ray strength functions for $^{207,208}$Pb as a function of $\gamma$-ray energy (see caption of Fig.~\ref{fig:model2}).}
\label{fig:model1}
\end {figure}

The contribution from isoscalar $E2$ transition strength is also included in the total $\gamma$-ray strength function and is described by~\cite{RIPL}
\begin{equation}
f_{E2}(E_\gamma) = \frac {1} {5 \pi^2 \hbar^2 c^2 E_\gamma^2} \frac {\sigma_{E2} E_\gamma \Gamma_{E2}^2}{(E_\gamma^2 -E_{E2}^2)^2 + E_\gamma^2 \Gamma_{E2}^2}.  
\label {eqn:23}
\end{equation}
The total model $\gamma$-ray strength functions, shown by solid lines in Figs.~\ref{fig:model2} and~\ref{fig:model1}, are the sum of the EGLO $E1$, Lorentzian $M1$ and $E2$ contributions, i.e.,
\begin{equation}
f_{\mathrm{tot}} = f_{\mathrm{EGLO}} + f_{M1} + E_{\gamma}^2f_{E2}. \\
\label {eq:24}
\end{equation}

The GEDR and GMDR parameters for $^{205-208}$Pb are taken from~\cite{RIPL} and are listed in Table II. The extracted normalized $\gamma$-ray strength functions of $^{205-208}$Pb are plotted together with the models discussed above in Figs.~\ref{fig:model2} and~\ref{fig:model1}. The photonuclear cross-section data of Ref.~\cite{yadfys} have also been drawn for comparison. 

The intermediate structures at the tail of GEDR in the Oslo data are observed at different $\gamma$-ray energies in the analyzed Pb nuclei. These intermediate structures present in our data oscillate between the above-mentioned models; however, none of them describe our data adequately for the whole energy region. In Figs.~\ref{fig:model2} and~\ref{fig:model1} our extracted data points (filled squares) of $^{205,206}$Pb and $^{207,208}$Pb tend to follow the EGLO model for $\gamma$-ray energies $\ge$ 4 MeV. For lower $\gamma$-energies the deviation between the theory and data points is obvious. This is not surprising, since one expects that the closed shell(s) will strongly influence the $\gamma$-decay, and thus preventing a good description of the $\gamma$-strength with smooth functions such as given by the EGLO model. 

The strength functions in $^{206-208}$Pb show an increase at $\gamma$-ray energies lower than $\sim$3 MeV. This may be due to the possible presence of strong non-statistical transitions, which are not correctly subtracted in the primary $\gamma$-ray extraction procedure and thus affecting the strength function at these $\gamma$-energies. Therefore, the enhanced $\gamma$-ray strength functions at low $\gamma$-ray energies are not conclusive.

\section{Summary and conclusions} 

The primary motivation of this work was to determine the level density and the $\gamma$-ray strength function in Pb isotopes near and at shell closure. The applicability of the Oslo method has also been investigated in the doubly magic $^{208}$Pb nucleus and its neighboring $^{205-207}$Pb nuclei. In contrast to the rare earth and mid-shell nuclei, these isotopes have low level density so that one can expect strong non-statistical fluctuations of level density and $\gamma$-ray strength function. 

The level densities and $\gamma$-ray strength functions of $^{205-208}$Pb have been extracted simultaneously from the primary $\gamma$-ray spectra. The comparison of our extracted level densities with spectroscopic measurements at low excitation energies gives good agreement within the experimental uncertainty. In $^{208}$Pb, the Oslo method could not give more information on level density than previously known from discrete spectroscopy. However, the good agreement between our data and known levels in $^{208}$Pb indicates the robustness of the method for its use in closed shell nuclei.

The level densities of $^{207,208}$Pb show significant step structures, which is an interesting finding of this work. Such structures are expected, partly due to strong shell effects at the $Z = 82$ and $N = 126$ shell closures, and partly due to the breaking of Cooper pairs. In $^{205,206}$Pb, these step structures are smoothed out as neutron valence holes come into play.

From the extracted level densities the micro-canonical entropies of the respective nuclei are deduced. The average entropy difference $\Delta S$ between $^{205}$Pb and $^{206}$Pb is found to be 0.9; however, the entropy difference between $^{207}$Pb and $^{208}$Pb varies violently with excitation energy and thus is difficult to use for finding other thermodynamic properties. The fluctuations in the entropy spectra are strongly enhanced in the temperature spectra. The average temperatures of the $^{205,206}$Pb nuclei are found to be $T \approx 1.0$ MeV. This is twice the temperature measured in rare earth nuclei.

The $\gamma$-ray strength functions of $^{205-208}$Pb show pronounced structures at energies below the neutron separation energy. However, the $\gamma$-ray strength function becomes smoother by the gradual opening of the neutron shell closure at $N=126$. Such structures have also been observed above the neutron threshold for ($\gamma$,n) reactions in $^{206-208}$Pb nuclei. The multipolarities of these intermediate resonances are unknown. The measured $\gamma$-ray strength functions of $^{205-208}$Pb are poorly described by the SLO and EGLO models in the energy region considered here. This indicates that there is more interesting physics connected to the shell closure(s), which is revealed in the $\gamma$-ray strength functions of the Pb isotopes.
\acknowledgments
Financial support from the Norwegian Research Council (NFR) is gratefully acknowledged.

\onecolumngrid

\begin{table}
\caption{Parameters used for the Fermi gas level density}

\begin{tabular}{l|cccccccc}
\hline
Nucleus		& $S_n$ & $a$		& $E_{pair}$	 &	$D_0$		&	$\rho_0(S_n)$	&	$D_1$ &	$\rho_1(S_n)$	&$\eta$	\\ 
			& (MeV) & (MeV$^-1$) & (MeV) & (keV)	&	 (10$^3$ MeV$^{-1}$)	&	(keV)	&	 (10$^3$ MeV$^{-1}$)& \\
\hline\hline
&&&&&\\

$^{205}$Pb & 6.732 & 15.03 & 0.878 & 2.0(5) & 29.4(78) & 0.8(2) & 25.4(68) & 0.30\\
$^{206}$Pb & 8.087 & 12.40$^{a}$ & 1.811 & - & - &- & 28(22)$^{b}$ & -\\
$^{207}$Pb & 6.738 & 11.15 & 1.005 & 32(6) &  1.58(34) & 5.9(9) & 2.98(55) & 0.46\\
$^{208}$Pb & 7.368 & 10.33 & 2.226 & 38(8) & 0.66(15) & 5(1) & 2.21(50) & 1.15\\
\hline\hline
\end{tabular}

$^{a}$ Determined using the empirical formula of the Gilbert and Cameron approach given in Ref.~\cite{RIPL} and references within.\\
$^{b}$ Estimated from systematics (see text of Fig.~\ref{fig:estimation}). \\
\label{tab:tab1}
\end{table}
\begin{table}
\caption{GEDR and GMDR parameters, used for the EGLO and SLO model calculations.}
\begin{tabular}{l|ccc|ccc|c}
\hline
Nucleus & $E_{{\mathrm{E}}1}$ & $\Gamma_{{\mathrm{E}}1}$ & $\sigma_{{\mathrm{E}}1}$ & $E_{{\mathrm{M}}1}$ & $\Gamma_{{\mathrm{M}}1}$ & $\sigma_{{\mathrm{M}}1}$ & $\langle \Gamma_\gamma \rangle$\\ 
 & (MeV) & (MeV) & (mb) & (MeV) & (MeV) & (mb) &(meV)\\
\hline\hline\\
$^{205}$Pb$\dagger$ & 13.59 & 3.85 & 514 & 6.95 & 4.0 & 1.16 & 330\\
$^{206}$Pb & 13.59 & 3.85 &  514 & 6.94 & 4.0 & 1.16 & - \\
$^{207}$Pb & 13.56 & 3.96 & 481 & 6.93 & 4.0 & 1.16 & 455(50) \\
$^{208}$Pb & 13.43 & 4.07 & 639 & 6.92 & 4.0 & 1.72 & - \\
\hline\hline

\end{tabular} 

$^{\dag}$ GEDR parameters taken from $^{206}$Pb.\\
\label{tab:tab2}

\end{table}

\newpage

\newpage

\newpage

\newpage

\newpage
\clearpage

\newpage

\newpage

\newpage

\newpage

\newpage

\newpage

\newpage

\newpage

\newpage

\newpage

\newpage

\newpage

\end{document}